\documentclass[aps,prb,reprint,superscriptaddress,showpacs]{revtex4-1}

\usepackage{graphics} 
\usepackage{amsmath} 
\usepackage{graphicx} 
\usepackage{dcolumn} 
\usepackage{bm} 
\usepackage{color}
\usepackage{subfigure} 
\newcommand{\rhoxo}{\rho_{x_{1}x_{1}}}
\newcommand{\rhoxt}{\rho_{x_{2}x_{2}}}
\newcommand{\rhob}{\rho_{bb}}
\newcommand{\rhobe}{\rho_{\beta\beta}}
\newcommand{\rhoal}{\rho_{\alpha\alpha}}
\newcommand{\gx}{\gamma_{x}}
\newcommand{\gh}{\gamma_{h}}
\newcommand{\gho}{\gamma_{1h}}
\newcommand{\ght}{\gamma_{2h}}
\newcommand{\gco}{\gamma_{1c}}
\newcommand{\gct}{\gamma_{2c}}
\newcommand{\Gc}{\Gamma_{c}}
 
\begin{document}
%\pagecolor{light-gray}
\title{An efficient biologically-inspired photocell enhanced by quantum coherence}
\author{C. Creatore} 
\affiliation{Cavendish Laboratory, University of Cambridge, Cambridge CB3 0HE, United Kingdom}
\author{M. A. Parker}
\affiliation{Cavendish Laboratory, University of Cambridge, Cambridge CB3 0HE, United Kingdom}
\author{S.  Emmott}
\affiliation{Microsoft Research, Cambridge CB1 2FB, United Kingdom}
\author{A. W. Chin }
\affiliation{Cavendish Laboratory, University of Cambridge, Cambridge CB3 0HE, United Kingdom}

\begin{abstract}
Artificially reproducing the biological light reactions responsible for the
remarkably efficient photon-to-charge conversion in photosynthetic complexes
represents a new direction for the future development of photovoltaic devices.
Here, we develop such a paradigm and present a model photocell based on the
nanoscale architecture of photosynthetic reaction centres that explicitly
harnesses the quantum mechanical effects recently discovered in photosynthetic
complexes. Quantum interference of photon absorption/emission induced by the
dipole-dipole interaction between molecular excited states guarantees an
enhanced light-to-current conversion and power generation for a wide range of
realistic parameters, opening a promising new route for designing artificial
light-harvesting devices inspired by biological photosynthesis and quantum
technologies.
\end{abstract}
%\date{\today}
%\pacs{42.50.Hz, 78.67.Hc, 03.65.Ud, 03.67.Lx}%
\maketitle
{\it Introduction.} Photosynthesis begins with an ultrafast
sequence of photo-physical events that convert solar photons into electrons for
use in the later dark stages of the process. Remarkably, in plants, bacteria
and algae, the photon-to-charge conversion efficiency of these light reactions
can approach $100\%$ under certain conditions\cite{blankenship}. This suggests a
careful minimisation of the deleterious molecular processes (trapping,
radiative and non-radiative losses, etc.) which plague attempts to achieve 
similar performances in artificial solar cells~\cite{wurfel}. Consequently, there has been
long-standing and ever-increasing interest in understanding the physics of the
nanoscale structures known as pigment-protein complexes (PPCs), 
which Nature uses to drive the light reactions efficiently~\cite{rvangrondellebook}. 
The subtlety and sophistication of PPCs have recently been underlined by the
unexpected observation of coherent quantum dynamics in several distinct types
of PPC~\cite{engel2007,calhoun2009,panit2010,hayes2010,collini2010}, raising
the intriguing idea that quantum effects could in fact contribute to the high
functional efficiency of PPCs. Demonstrating a link between efficiency,
function and the stabilisation of room-temperature quantum effects in these
nanoscale system could have a significant impact on the design of future
quantum-based nanotechnologies. 

Recently, Dorfman {\it et al.}~\cite{scully2013} have introduced a promising 
approach to this problem, in which the light reactions are analysed as quantum heat engines
(QHEs). Treating the light-to-charge conversion as a continuous Carnot-like
cycle, Dorfman {\it et al.} indicate that quantum coherence could boost
the photocurrent of a photocell based on photosynthetic reaction centres by at
least $27\%$ compared to an equivalent classical photocell. In their model the driver of this
enhancement is the phenomenon of Fano interference \cite{scully2011a,scully2011b,scully2013}, 
which has been demonstrated to enable optical systems to violate the thermodynamic
detailed balance that otherwise limits the efficiency of light-harvesting
devices, as shown by Shockley and Queisser in their celebrated calculation of
the fundamental limit of semiconductor solar cells \cite{shockley1961}. 
\begin{figure}[t!]
\centering
\includegraphics[width=0.45\textwidth]{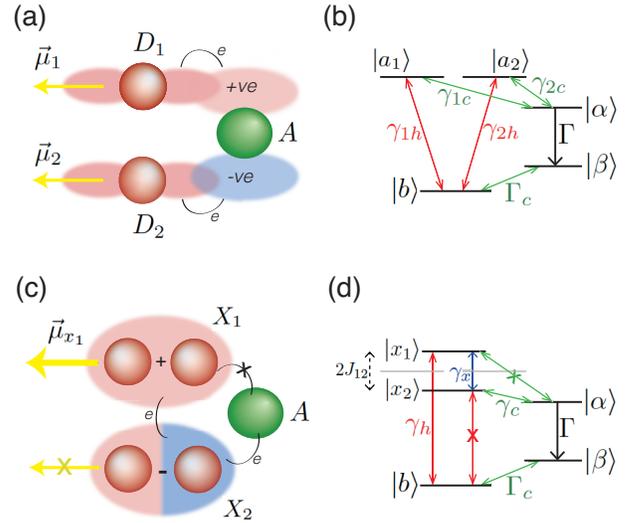}%model.eps
\caption{(color online). Schematic of the reaction centre. In (a), both 
donors $D_{1}$ and $D_{2}$ are optically active and cooperatively contribute to transferring the excited electrons at the acceptor $A$. 
The red and blue shadowed/pale regions surrounding the molecules denote the molecular orbitals representing the spatial distribution density of electrons. 
In (c), coupling between $D_{1}$ and $D_{2}$ gives rise to a coupled system ($X_{1}\,,X_{2}$) with new eigenstates resulting 
from the symmetric \textit{bright} ($|x_{1}\rangle$) and antisymmetric \textit{dark} ($|x_{2}\rangle$) superpositions of the uncoupled donor states 
($|a_{1}\rangle\,,|a_{2}\rangle$). Photon emission/absorption is only possible via $|x_{1}\rangle$, while charge transfer to $A$ occurs via $|x_{2}\rangle$ only. 
(b) and (d) show the level schemes of  (a) and (c), respectively.
}
\label{fig:schematic}
\end{figure}

Here we develop the idea of photosynthesis-inspired structures as QHEs, showing
that naturally-occurring dipole-dipole interactions between molecular pigments
in PPCs generate quantum interference effects that can enhance the photocurrents and
maximum power outputs by $>30\%$. Our device \emph{avoids} the need
to invoke Fano interference, which is a higher-order perturbative interaction, whose 
mathematical description in [\onlinecite{scully2013}] appears to suffer from a number of pathologies 
including the violation of positivity of the quantum master equation, as 
documented in the Supplemental Material~\cite{sm}. In this paper, we set out the electronic 
structure and spatial nanostructure required to design a simple and more robust device, 
which is based on a completely positive master equation, uses direct (first order) interactions 
and could be easily interfaced with other emerging nanotechnologies inspired by biological 
light-harvesting structures \cite{higgins2013}.

{\it Model.} The cyclic engine model we propose [see Figs.1(a)-(b)] mimics the `special pair' in photosynthetic 
RCs as in the original scheme by Dorfman \cite{scully2013}: 
$D_{1}$ and $D_{2}$ represent a pair of identical and initially
uncoupled donor molecules which flank an acceptor molecule $A$. The
cycle begins with the absorption of solar photons - the whole system being initially in 
the ground state $|b\rangle$ - leading to the population of the donor excited states $|a_{1}\rangle$ 
and $|a_{2}\rangle$. The excited electrons can then be transferred to the acceptor molecule 
(electrons being in $A$ and holes remaining in $D_{1},\,D_{2}$) through electronic coupling and
emission of phonons, as in [\onlinecite{scully2013}]. Then, the charge separated state
$|\alpha\rangle$ decays to a state $|\beta\rangle$ representing
the now positively charged RC. The
transfer rate $\Gamma$ and steady ratio of populations between $|\alpha\rangle$ and
$|\beta\rangle$ determine the current $j=e\Gamma\rho_{\alpha\alpha}$ and power output of our QHE, 
$\rho_{\alpha\alpha}$ being the population of $|\alpha\rangle$. Finally,
the cycle is closed allowing the $|\beta\rangle$ state to decay back (via 
a rate $\Gamma_{c}$) to the neutral ground state of the system as the $|\beta\rangle$ state is reduced
by other parts of the photosynthetic apparatus\cite{blankenship}.

The new element of our scheme is the formation of new optically excitable
states through strong excitonic coupling between the donors,
resulting from their long-range optical transition dipole-dipole interaction~\cite{sm}. 
Such interactions and the
formation of stable delocalised excited (exciton)
states are widely observed phenomena in pigment-protein
complexes~\cite{rvangrondellebook}, and provide the quantum interference
effects required to enhance the photocurrent and power of our QHE. 
Following the standard interpretation of most experimental studies of PPCs, 
we assume that the new donor states induced by the excitonic coupling 
are the states which get populated by the absorption of photons. 
Our QHE scheme will therefore only consider 
the populations of these new states in the work cycle, and the effects of quantum coherence 
and interference, vide infra, will be included in the coherent modification of the probability 
amplitudes for light absorption, relaxation and electron transfer for the exciton states.

Within our model we consider identical and degenerate donor excited states with parallel 
dipole moments $|\vec{\mu}_{1}|=|\vec{\mu}_{2}|=\mu$ and electron transfer matrix elements 
$|t_{D_{1}A}|=|t_{D_{2}A}|=t_{DA}$ resulting from the overlap between the donor's and acceptor's 
electron wavefunctions. Further, we exploit the possibility of using acceptors molecules, which can host an electron 
into a lowest unoccupied molecular orbital with a spatially varying phase. For an acceptor with orbital lobes characterised 
by a relative phase of $\pi$ [see Fig.~1(a)], placing the donor molecules close to different lobes, 
leads to electron transfer matrix elements with the same magnitude (same overlap), 
but opposite signs, i.e. $t_{D_{2}A}=-t_{D_{1}A}$. The new donor eigenstates in the presence 
of the dipolar excitonic coupling $J_{12}$
%$H_{I}=J_{12}(|a_{1}\rangle\langle a_{2}| + (|a_{2}\rangle\langle a_{1}|)$ 
are symmetric ($|x_{1}\rangle$) and antisymmetric ($|x_{2}\rangle$) combinations of the uncoupled donor states: 
$\displaystyle |x_{1}\rangle=1/\sqrt{2}(|a_{1}\rangle + |a_{2}\rangle)$ and $|x_{2}\rangle=1/\sqrt{2}(|a_{1}\rangle - |a_{2}\rangle)$, 
with corresponding energy eigenvalues $\displaystyle E_{x_{1}/x_{2}}=E_{1,2} \pm  J_{12}$, $E_{1,2}$ being the energies 
of the uncoupled states $|a_{1}\rangle$ and $|a_{2}\rangle$, respectively~\cite{sm}. 
The dipole moment of $|x_{1}\rangle$ is therefore enhanced by \emph{constructive}
interference of the individual transition dipole matrix elements, $\mu_{x_{1}}=1/\sqrt{2}(\mu_{1}+\mu_{2})=\sqrt{2}\mu$, 
whereas the dipole moment of $|x_{2}\rangle$ cancels due to \emph{destructive} interference. 
Hence, the symmetric combination describes an optically active \textit{bright} state with a photon absorption/emission rate 
$\gamma_{h}\propto |\mu_{x_{1}}|^2=2|\mu|^{2}$, which is twice that of the uncoupled donor states ($\gamma_{1h}=\gamma_{2h}\propto\,|\mu|^2$), 
while the antisymmetric combination describes an optically forbidden \textit{dark} state. 
Crucially for our scheme, the bright state lies $2J_{12}$ higher in energy compared to the dark state. 
Similarly, for the donor-acceptor arrangement of Fig.~1, the probability amplitudes for
electron transfer coherently interfere such that the bright state has a
resultant matrix element equal to zero and charge transfer to the acceptor $A$
occurs via the dark state only, with a matrix element
$t_{x_{2}A}=\sqrt{2}t_{DA}$ and a electron transfer rate $\gamma_{c}\propto\,|t_{x_{2}A}|^2=2|t_{DA}|^2$, 
i.e. twice the rate of the uncoupled donors, $\gamma_{1c}=\gamma_{2c}\propto\,|t_{DA}|^2$.

With this scheme the photon absorption and electron transfer parts of the cycle are disconnected unless there is population transfer between the bright and dark states. 
This is provided by the phonon-mediated energy relaxation, which can be very effective in excitonic systems~\cite{rvangrondellebook} and 
is included in our kinetic model via the relaxation rate $\gamma_{x}$ [Fig.~1(d)]. 
Treating the donor-light, electron transfer and bright-dark relaxation couplings in 2$^{\textrm{nd}}$ order perturbation theory, 
the kinetics of the photo-induced charge transfer described in Figs.~1(a),1(c), using the level schemes of Figs.~1(b),1(d), 
obeys the following Pauli master equation, which is guaranteed to give completely positive populations: 
 \begin{equation}
\begin{split}
\label{eq:rhoii}
\dot{\rho}_{a_{1}a_{1}}  = &  -\gho[(1+n_{1h})\rho_{a_{1}a_{1}} -n_{1h}\rhob]\\
&-\gco[(1+n_{1c})\rho_{a_{1}a_{1}}-n_{1c}\rhoal]\,, \\
\dot{\rho}_{a_{2}a_{2}} = &  -\ght[(1+n_{2h})\rho_{a_{2}a_{2}} -n_{2h}\rhob]\\
& -\gct[(1+n_{2c})\rho_{a_{2}a_{2}}-n_{2c}\rhoal]\,,
\end{split}
\end{equation}
\begin{equation}
\begin{split}
\label{eq:rhoxixi}
\dot{\rho}_{x_{1}x_{1}} = & -\gx[(1+n_{x})\rho_{x_{1}x_{1}} -n_{x}\rho_{x_{2}x_{2}}]\\
&-\gh[(1+n_{h})\rho_{x_{1}x_{1}} -n_{h}\rhob]\,,\\
\dot{\rho}_{x_{2}x_{2}} = &\,\, \gx[(1+n_{x})\rho_{x_{1}x_{1}} -n_{x}\rho_{x_{2}x_{2}}]\\
&-\gamma_{c}[(1+n_{2c})\rho_{x_{2}x_{2}}-n_{2c}\rhoal]\,.
\end{split}
\end{equation}
In Eqs.~(1)-(2), $n_{1h}$ ($n_{2h}$) and $n_{h}$ are the average numbers of photons with frequencies matching the transition energies from the ground state to $|a_{1}\rangle$ ($|a_{2}\rangle$) and $|x_{1}\rangle$, respectively; 
$n_{1c}$ ($n_{2c}$) are the thermal occupation numbers of ambient phonons at temperature $T_{a}$ with energies $E_{1}-E_{\alpha}$ 
($E_{2}-E_{\alpha}$) for $J_{12}=0$ [Eq.~(1)] and $E_{x_{1}}-E_{\alpha}$ 
($E_{x_{2}}-E_{\alpha}$) for $J_{12}\neq\,0$ [Eq.~(2)]; $n_{x}$ is the corresponding thermal occupation at $T_{a}$ with 
energy $E_{x_{1}}-E_{x_{2}}$. The rates in Eqs.~(1)-(2) all obey local detailed balance, and correctly lead to a Boltzmann distribution for the level populations if the thermal averages for the photon and phonon reservoirs are set to a common temperature. We consider the fully populated ground state, i.e. $\rho_{bb}(t=0)=1$, 
as the initial condition. The full quantum master equation (QME) describing the evolution of all the populations of the states in the QHE cycle is given 
in the SM text~\cite{sm}.

\begin{table}[t!]
%\caption{}
\centering
\begin{tabular}{|c| c| c|}
\hline
%%%
&Energies and& Occupation\\
&Decay rates (eV)& numbers\\
\hline
E$_{1}-E_{b}$ &  1.8 &\\
E$_{2}-E_{b}$ &  1.8 &\\ 
E$_{1}$-E$_{\alpha}$  & 0.2&\\
E$_{2}$-E$_{\alpha}$   & 0.2&\\
E$_{\beta}$-E$_{b}$  &  0.2&\\
$J_{12}$ &  0.015&\\
%$T_{a}$ & & 300K&\\
$\gamma_{h}$ & 1.24$\times$ 10$^{-6}$&\\
$\gamma_{1h}=\gamma_{2h}$&  0.62$\times$ 10$^{-6}$&\\
$\Gamma$ &  0.124&\\
$\Gamma_{c}$ &  0.0248&\\
$n_{x}$&  &0.46\\
$n_{h}=n_{1h}=n_{2h}$& &60000\\ 
\hline
\end{tabular}
\caption{Model parameters used in the numerical simulations.}
\label{tbl:t1}
\end{table}

{\it Results.} The steady state solutions of the QME~\cite{sm} can be used to evaluate the current generated in our model RC, and thus the current enhancement induced by the coherent coupling with respect to the same scheme with uncoupled donors. Figure 2 shows 
the relative enhancement $(j-\tilde{j})/\tilde{j}$, $j$ and $\tilde{j}$ being the electric current in the coupled ($J_{12}\neq\,0$) and uncoupled ($J_{12}=0$) case, respectively. The current enhancement has been evaluated at room temperature as a function of the transition rates $\gamma_{x}$ and $\gamma_{c}$ ($=\gamma_{1c}+\gamma_{2c}$ 
with $\gamma_{1c}=\gamma_{2c}$) using realistic photon decay rates and average numbers of solar concentrated photons and ambient phonons as listed in Table~\ref{tbl:t1} and reported in  [\onlinecite{scully2013}]. When $\gamma_{x}=\gamma_{c}$ (see the continuous red line), 
the steady-state currents yield the same value ($j=\tilde{j}$): charge-transfer via the channel $x_{1}\rightarrow\,x_{2}\rightarrow\,\alpha$ is as fast as the total combined transfer through the independent channels $a_{1}\rightarrow\,\alpha$ and $a_{2}\rightarrow\,\alpha$. However, when $\gamma_{x}>\gamma_{c}=\gamma_{1c}+\gamma_{2c}$, coherent coupling  leads to substantial current enhancements as compared to the configuration without coupling. These enhancements increase monotonically with increasing 
$\gamma_{x}$ when $\gamma_{x}>\gamma_{c}$. These results can be understood as arising from the new delocalised level structure, which introduces an effective shelving state, i.e. the dark state, into the cycle. The advantage of having a dark shelving state is that the deleterious emission process are efficiently outcompeted by the 
fast {\it bright-to-dark} relaxation after the light absorption. Further, when $\gamma_{x}>\gamma_{c}$, the absorbed solar energy can be removed faster than the absorbers in the incoherent case can move it to the acceptor $|\alpha\rangle$ state.  Then, as emission losses are absent in the dark state, all its population must pass through the work (extracting) stage. 
\begin{figure}[t!]
\begin{center}
\includegraphics[width=0.4\textwidth]{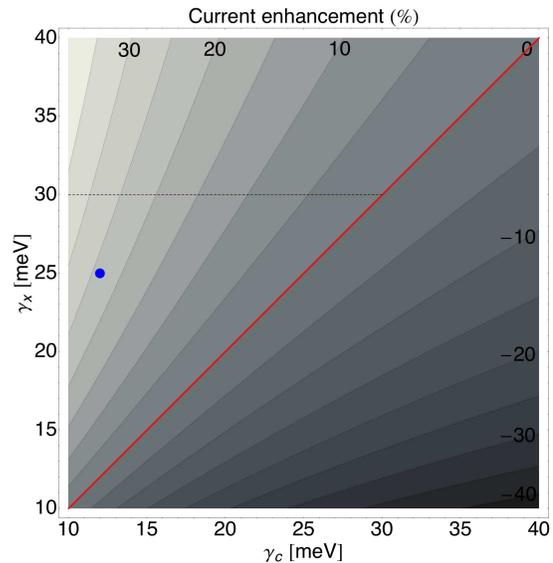}
\end{center}
\caption{(color online). Contour plot of the percentage current enhancement $(j-\tilde{j})/\tilde{j}$ 
induced by coherent coupling between the donor molecules and calculated as a function of the phonon 
decay rates $\gamma_{x}$ and $\gamma_{c}$ at  $T_{a}=300$K.}
\label{fig:enhanceEV}
\end{figure}
Hence, we have a type of controllable \textit{valve} through which we can extract solar energy fast and use it to build up higher voltages/driving forces to extract power in the latter part of the cycle~\cite{miller2012}. 

We have assumed that the delocalised states formed through the dipolar coupling are stable under steady state operation, whereas dephasing of the these states 
due to energy relaxation, may suppress their phase coherence and the interference of transitions previously discussed. While a sophisticated treatment of the non-equilibrium open system dynamics of our scheme is beyond the scope of this article, we employ the standard condition that the coherent superposition states will be stable provided the energy splitting $2J_{12}>\gamma_{x}$~\cite{schon2002}.
This condition limits the achievable enhancement in conditions related to realistic RCs: 
for the physical parameters of Table \ref{tbl:t1}, the enhancement is restricted to $~30\%$ (see the dashed black line $\gamma_{x}=2J_{12}=30$ meV). 

We note that we have taken an excitonic coupling strength corresponding to a splitting of $\approx 240 \mathrm{cm}^{-1}$, which is actually towards the lower end of the typical excitonic splittings in natural RCs ($200-800 \mathrm{cm}^{-1}$) \cite{scully2013}. Hence, this enhancement is likely to be highly robust, or even larger, in schemes employing couplings realisable in natural systems. Moreover, these larger splittings, which can be much larger than thermal energy at room temperature ($k_{b}T_{a}\approx210 \mathrm{cm}^{-1}$), lead to other advantages which we now discuss. 
\begin{figure}[t!]
\begin{center}
\includegraphics[width=0.4\textwidth]{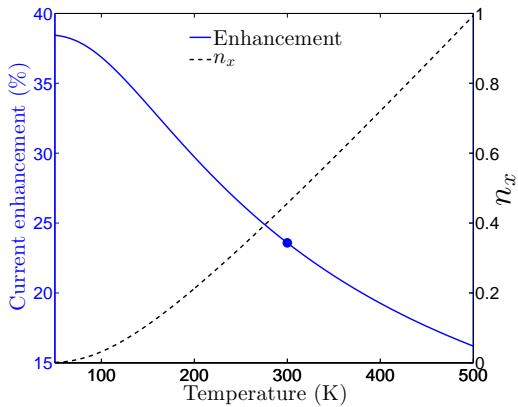}
\end{center}
\caption{(color online). Temperature dependence of the current enhancement. The enhancement $(j-\tilde{j})/\tilde{j}$ 
and average number of phonons $n_{x}$ evaluated as a function of temperature at fixed phonon rates $\gamma_{c}=12$ meV and $\gamma_{c}=25$ meV
(see the blue dot in Fig.2).}
\label{fig:enhanceT}
\end{figure}

Figure~3 shows how the current enhancement changes as function of temperature for fixed transition rates and dipolar coupling. As the temperature decreases, so does the average number of phonons $n_{x}$, reducing the thermal uphill transition of the dark to bright state; under these conditions the transition from the bright to the dark state ($x_{1}\rightarrow\,x_{2}$) become essentially unidirectional and the spontaneous emission losses from the bright to ground state ($x_{1}\rightarrow\,b$) are further reduced, thus resulting in a more efficient charge transfer ($x_{2}\rightarrow\,\alpha$) at the acceptor. 

Within this scheme, the current generated can be thought to flow across a {\it load} 
connecting the acceptor levels $\alpha$ and $\beta$. The voltage $V$ across this load can be expressed as 
$eV=E_{\alpha}-E_{\beta} + k_{b}T_{a}\textrm{ln}(\rhoal/\rhobe)$, $e$ being the electric charge~\cite{scully2013,miller2012,wurfel}. 
The performance of our RC-inspired QHE can be thus assessed in terms of its photovoltaic properties, 
calculating the steady-state current-voltage ($j-V$) characteristic and power generated. 
The $j-V$ characteristic and power $P=j\cdot\,V$ are evaluated using the steady state solutions of the QME, 
calculated at increasing rate $\Gamma$ at fixed other parameters: from $\Gamma\rightarrow\,0$ ($j\rightarrow\,0$), corresponding to 
the \textit{open circuit} regime where $eV=eV_{oc}=(E_{a_{1}/x_{1}}-E_{b})(1-T_{a}/T_{S})\approx\,E_{a_{1}/x_{1}}-E_{b}$, 
to the \textit{short circuit} regime where $V\rightarrow\,0$. 
\begin{figure}[ht!]
\begin{center}
\includegraphics[width=0.45\textwidth]{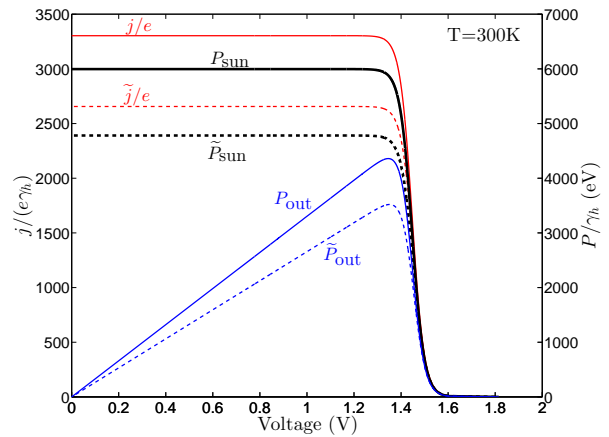}\\
\end{center}
\caption{(color online). Current-voltage characteristic and power generated at room temperature. The red thin continuous and dashed lines 
represent the current-voltage dependence in the presence ($j/e$, $J_{12}\neq\,0$) and 
absence ($\tilde{j}/e$, $J_{12}=0$) of coherent coupling, respectively. The power generated by the photosynthetic reaction centre 
as a function of the induced cell voltage $V$ is indicated by blue thin continuous ($P_{\textrm{out}}$, $J_{12}\neq\,0$) 
and dashed ($\widetilde{P}_{\textrm{out}}$, $J_{12}=0$) lines. The power acquired from the sun is $\widetilde{P}_{\textrm{sun}}=\tilde{j}\cdot(E_{a_{1}}-E_{b})/e$ for $J_{12}=0$ 
(thick black dashed line) and $P_{\textrm{sun}}=j\cdot(E_{x_{1}}-E_{b})/e$ for $J_{12}\neq\,0$ (thick continuous black  line).}
\label{fig:vjp}
\end{figure}
The $j-V$  and the $P-V$ behaviours are shown in Fig.~4 
for both cases of coupled ($J_{12}\neq\,0$) and uncoupled ($J_{12}=0$) donor molecules. Using the parameters listed in Table~\ref{tbl:t1} 
with $\gamma_{c}=12$ meV and $\gamma_{x}=25$ meV (see also Fig.~2), we get a peak current enhancement $\approx$ 24\%. 
The enhancement of the delivered peak power in the presence of coherent coupling is quantified defining 
the relative efficiency $\eta_{R}=(P_{out}^{\textrm{max}}-\tilde{P}_{out}^{\textrm{max}})/\tilde{P}_{out}^{\textrm{max}}$, 
$P_{out}^{\textrm{max}}$ and $\tilde{P}_{out}^{\textrm{max}}$ being the output peak powers with ($J_{12}\neq\,0$) 
and without ($J_{12}=0$) coupling, respectively. For the parameters used in the simulation shown in Fig.~4, 
$\eta_{R}\approx$ 24\%.
The coherent RC-based photocell thus generates higher peak powers by drawing a larger current/energy flux from the same available solar 
radiation source whilst maintaining the same voltage. 
Furthermore, for decreasing temperatures the efficiency of charge transfer increases and thus improved photocell performances 
with relative efficiencies $\eta_{R}$ up to 40\% can be achieved (see SM).

{\it Conclusions.} We have shown that quantum interference effects resulting from the strong dipolar interactions 
naturally occurring in RCs pigments, can considerably enhance the performance of a QHE 
relative to an incoherent set up. 

The potential for harnessing quantum effects such as the formation of coherent super positions, 
has been highlighted in a number of previous works relating to the efficiency of energy transfer 
in PPCs \cite{plenio2008,mohseni2008,caruso2009,rebentrost2009,chin2010,chin2013,delrey2013}, 
and such states and quantum dynamics have been observed in both ensemble and single-molecule experiments on PPCs~\cite{engel2007,calhoun2009,panit2010,hayes2010,collini2010,hayes2013} and RCs~\cite{westenhoff2012}. 
This fascinating potential may soon be realised in
artificial molecular light-harvesting systems, such as those recently synthesised by Hayes {\it et al.} \cite{hayes2013}, 
although advanced simulation techniques \cite{novoderezhkin2004,prior2010,chinchain2010,chin2013,ishizaki2009,nalbach2011} 
will be required to assess the optimal design and stability of quantum states of the devices for PPC-like 
or related organic photovoltaic materials.

{\it Acknowledgements.} A.\,W.\,C. acknowledges support from the Winton
Programme for the Physics of Sustainability; C.\,C. acknowledges support from EPSRC. 
We would like to thank Nick Hine and Akshay Rao for useful discussions.

%\clearpage
%\bibliography{prova} 

\newpage
\clearpage
\onecolumngrid
\section*{Supplemental Material}
\subsection*{Coherently coupled molecules}
We consider two molecules modelled as two-level systems (TLSs) and 
coupled to the radiation field via dipole transitions.  We assume the TLSs to be described by their classical dipoles 
$\vec{\mu}_{1}=\mu_{1}\hat{\mu}_{1}$ and  $\vec{\mu}_{2}=\mu_{2}\hat{\mu}_{2}$ oscillating at frequencies $\omega_{1}=ck_{1}$ 
and $\omega_{2}=ck_{2}$, respectively (energies being $E_{1}=\hbar\omega_{1}$ and $E_{2}=\hbar\omega_{2}$), 
and generating electric fields at position $\vec{r}$, $\vec{E}_{1}(\vec{r})$ and $\vec{E}_{2}(\vec{r})$.
 The potential energy of the second dipole $\vec{\mu}_{2}$ at position $\vec{r}_{12}$ is modified 
 by the field of the first dipole and given by~\cite{jackson} $V_{12}=-\vec{\mu}_{2}\vec{E}_{1}(\vec{r}_{12})$. 
 Here $\vec{r}_{12}$ is the position of $\hat{\mu}_{2}$, relative to a frame centered in $\hat{\mu}_{1}$. 
 For dipoles oscillating with approximately the same frequency, $V_{12}\approx V_{21}\equiv V$. 
 The real part of $V$ represents the coupling $J_{12}$ and results in a 
 \textit{coherent} interaction in which no energy is emitted from the coupled system of the two TLSs into the radiation field. 
 If the distance between the TLSs is much smaller than the transition wavelength, i.e. $r_{12}\ll 1/k$, then 
 $\displaystyle J_{12}\propto\,(1/r_{12}^3)f(\hat{\mu}_{1},\hat{\mu}_{2})$, where
  $f(\hat{\mu}_{1},\hat{\mu}_{2})$ is a function of the mutual orientation between the dipoles. Hence, the dipole-dipole interaction 
 has a $1/r^{3}$ dependence and is influenced by the orientation of the dipoles. In the case of parallel dipoles next to each other,  
 as considered in this article, $f(\hat{\mu}_{1},\hat{\mu}_{2})=1$.  The Hamiltonian for the two TLSs interacting via 
 the dipole-dipole interaction is given by 
 \setcounter{equation}{0}
 \begin{equation}
\renewcommand{\theequation}{S\arabic{equation}} 
\label{eq:hamil}
 H=H_{0} + H_{I}\,,
 \end{equation}
 where $H_{0}=\sum_{i=1,2}\hbar\omega_{i}\sigma^{+}_{i}\sigma^{-}_{i}$ and 
 $H_{I}=J_{12}(\sigma_{1}^{-}\sigma_{2}^{+} + h.c.)$. $\sigma^{+}_{i}=|e_{i}\rangle\langle g_{i}|$ and 
 $\sigma^{-}_{i}=|g_{i}\rangle\langle e_{i}|$ are dipole raising and lowering operators, respectively with $e_{i}$ 
 ($g_{i}$) the excited (ground) state of the $i^{\textrm{th}}$ TLS. The interaction part $H_{I}$ accounts for the 
 conservation of the energy in the system, associated with the exchange of virtual photons between the two TLSs. 
 Diagonalisation of the Hamiltonian Eq.~(S1) yields the new eigenstates for the coupled molecules:
 $|xx\rangle=|e_{1}e_{2}\rangle$, $|x_{1}\rangle=\textrm{sin}\theta|e_{1}g_{2}\rangle +  \textrm{cos}\theta|e_{2}g_{1}\rangle$, 
 $|x_{2}\rangle=\textrm{cos}\theta|e_{1}g_{2}\rangle -  \textrm{sin}\theta|e_{2}g_{1}\rangle$, $|gg\rangle=|g_{1}g_{2}\rangle$, having 
 corresponding eigenvalues $E_{xx}=E_{1}+E_{2}$, $E_{x_{1}}=(E_{1}+E_{2})/2 + \sqrt{(E_{1}-E_{2})^{2}/4 + J_{12}^2}$, 
 $E_{x_{2}}=(E_{1}+E_{2})/2 - \sqrt{(E_{1}-E_{2})^{2}/4 + J_{12}^2}$, $E_{gg}=0$ and mixing angle $\theta$ given by 
 $\textrm{tan}(2\theta)=2J_{12}/(E_{1}-E_{2})$. Hence, for degenerate molecules ($E_{1}=E_{2}\implies\theta=\pi/4$) 
 (with parallel dipoles) the $|x_{1}\rangle$ and $|x_{2}\rangle$ eigenstates read: $|x_{1}\rangle=1/\sqrt{2}(|e_{1}g_{2}\rangle+
 |g_{1}e_{2}\rangle)=1/\sqrt{2}(|a_{1}\rangle+|a_{2}\rangle)$, and 
 $|x_{2}\rangle=1/\sqrt{2}(|e_{1}g_{2}\rangle-|g_{1}e_{2}\rangle)=1/\sqrt{2}(|a_{1}\rangle-|a_{2}\rangle)$.

\subsection*{Quantum Master Equations}
%\begin{widetext}
We consider the model of the reaction centre with uncoupled donor molecules [see Fig. 1(a)]. 
The evolution of the density matrix elements based on the application of detailed balance to the level scheme shown 
in Fig.~1(b) is given by the following quantum master equation (QME)
\begin{equation}
\renewcommand{\theequation}{S\arabic{equation}} 
\begin{array}{rcl}
\dot{\rho}_{a_{1}a_{1}}  &=& -\gho[(1+n_{1h})\rho_{a_{1}a_{1}} -n_{1h}\rhob]-\gco[(1+n_{1c})\rho_{a_{1}a_{1}}-n_{1c}\rhoal]\,, \\
\dot{\rho}_{a_{2}a_{2}}  &=&  -\ght[(1+n_{2h})\rho_{a_{2}a_{2}}-n_{2h}\rhob]-\gct[(1+n_{2c})\rho_{a_{2}a_{2}}-n_{2c}\rhoal]\,,   \\
\dot{\rho}_{\alpha\alpha} &=&  \gco[(1+n_{1c})\rho_{a_{1}a_{1}}-n_{1c}\rhoal] + \gct[(1+n_{2c})\rho_{a_{2}a_{2}}-n_{2c}\rhoal] - \Gamma\rhoal\,,  \\
\dot{\rho}_{\beta\beta}  &=&  \Gamma\rhoal -  \Gc[(1+N_{c})\rhobe - N_{c}\rho_{bb}]\,,    \\
\rho_{a_{1}a_{1}}  &+&  \rho_{a_{2}a_{2}}+ \rhoal + \rhobe + \rhob = 1\,.
\end{array}
\label{eq:qme_nint2}
\end{equation}
When the excited states ($a_{1}$, $a_{2}$) of the donor molecules are coherently coupled and new eigenstates $x_{1}$ and $x_{2}$ 
are formed [see Fig.~1(c) and the level scheme Fig.~1(d)], the QME takes the form
\begin{equation}
\renewcommand{\theequation}{S\arabic{equation}} 
\begin{array}{rcl}
\dot{\rho}_{x_{1}x_{1}} &=& -\gx[(1+n_{x})\rhoxo -n_{x}\rhoxt]-\gh[(1+n_{h})\rhoxo -n_{h}\rhob]\,,\\
\dot{\rho}_{x_{2}x_{2}} &=& \gx[(1+n_{x})\rhoxo -n_{x}\rhoxt]-\gamma_{c}[(1+n_{2c})\rhoxt-n_{2c}\rhoal]\,,  \\
\dot{\rho}_{\alpha\alpha} &=&  \gamma_{c}[(1+n_{2c})\rhoxt-n_{2c}\rhoal] - \Gamma\rhoal\,, \\
\dot{\rho}_{\beta\beta} &=& \Gamma\rhoal -  \Gc[(1+N_{c})\rhobe - N_{c}\rho_{bb}]\,,\\
\rhoxo &+& \rhoxt + \rhoal + \rhobe + \rhob = 1\,.
\end{array}
\label{eq:qme_int2}
\end{equation}
In Eqs.~(S2)-(S3), $n_{1c}$ ($n_{2c}$) are the average occupation numbers of phonons at 
ambient temperature $T_{\textrm{a}}$ with energies $\Delta E=E_{1}-E_{\alpha}$ ($E_{2}-E_{\alpha}$) for $J_{12}=0$ [Eq.~(S2)] 
and $\Delta E=E_{x_{1}}-E_{\alpha}$ ($E_{x_{2}}-E_{\alpha}$) for $J_{12}\neq\,0$ [Eq.~(S3)]; 
likewise $n_{x}$ and $N_{c}$ are the average thermal occupations at $T_{a}$ with energies $\Delta E=E_{x_{1}}-E_{x_{2}}$ and  
$\Delta E=E_{\beta}-E_{b}$, respectively. The average thermal occupations at energies $\Delta E$ are given according to the Planck distribution:
\begin{equation}
\renewcommand{\theequation}{S\arabic{equation}}  
n=\frac{1}{e^{\Delta E/(k_{B}T_{a})} -1} 
\end{equation}
Both quantum master equations Eqs.~(S2)-(S3) are guaranteed to yield always positive solutions for the populations. 
As explained in the main text, due the constructive interference between the dipole moments in the symmetric bright state ($x_{1}$), $\gamma_{1h}=\gamma_{2h}=\gamma_{h}/2$. Likewise, due to the constructive interference between the electric matrix elements in the dark state ($x_{2}$), 
$\gamma_{1c}=\gamma_{2c}=\gamma_{c}/2$. Figures~S1(a),S1(b) show the time evolution of the QME populations 
(density matrix elements) obtained solving Eqs.~(S2)-(S3). For this particular example we have used realistic values $\gamma_{x}\approx$ 25 meV, $\gamma_{c}$=12 meV, $\gamma_{h}=1.24\,\mu$eV, $\Gamma_{c}\approx$ 25 meV, 
$\Gamma=0.12$ eV, $n_{x}\approx$ 0.46 at $T_{a}$= 300K, and average numbers of solar concentrated photons $n_{h}=n_{1h}=n_{2h}=60000$.

\setcounter{figure}{0}
\begin{figure}[h!]
\renewcommand{\thefigure}{S\arabic{figure}}
\begin{center}
\subfigure{%
\label{fig:ex_nint}
\includegraphics[width=0.45\textwidth]{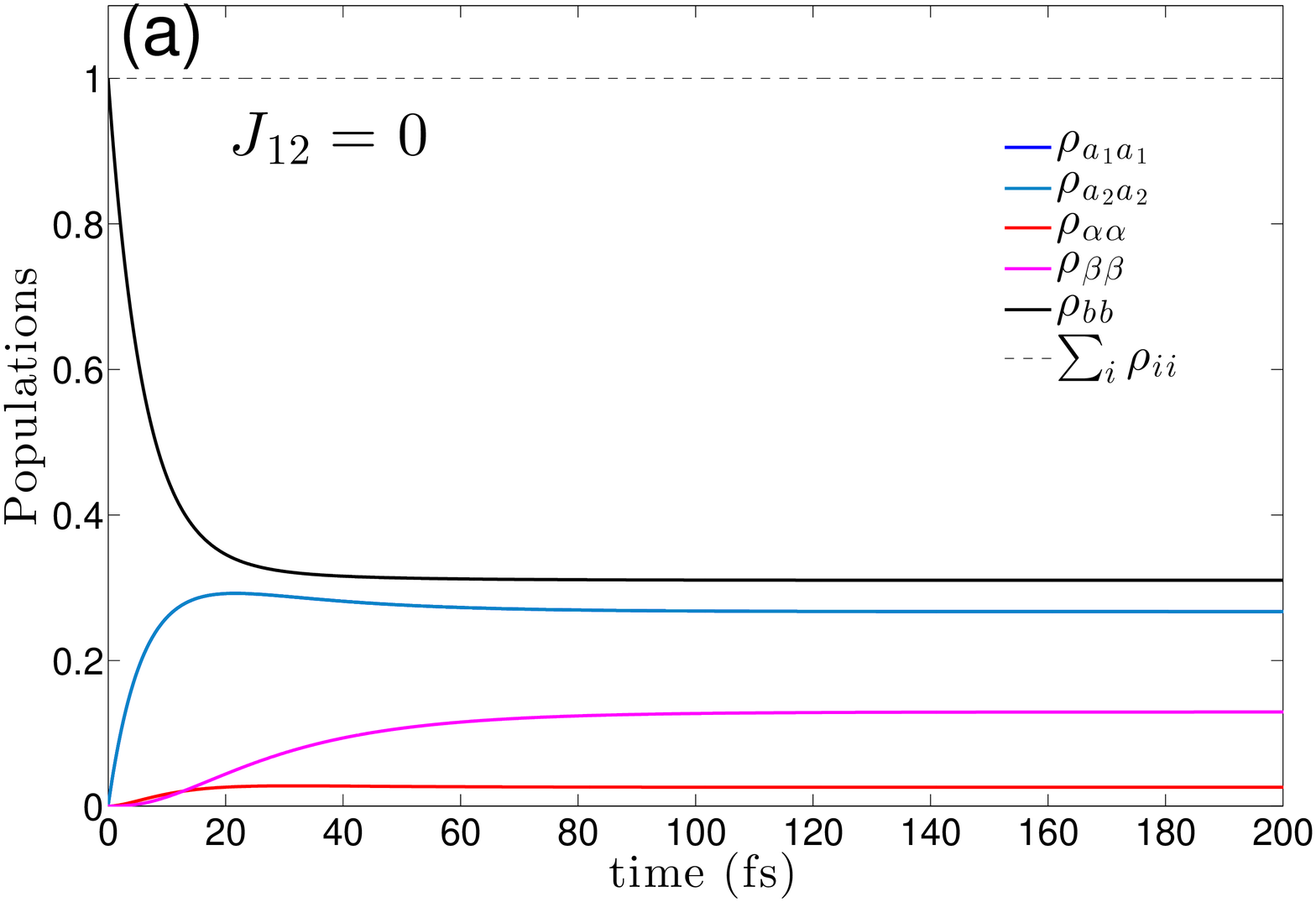} 
}%
\subfigure{%
\label{fig:ex_int}
\includegraphics[width=0.45\textwidth]{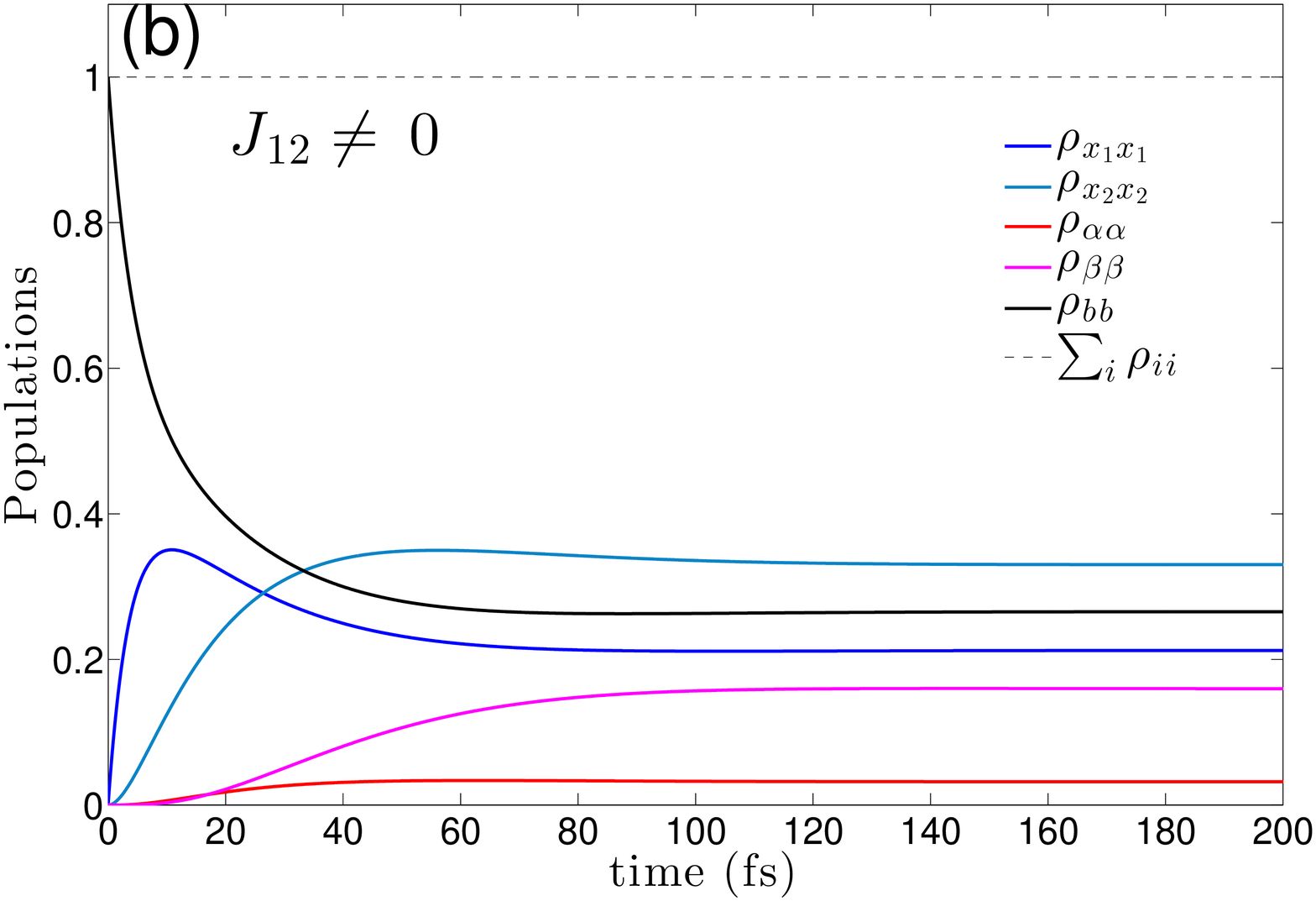} 
}%
\end{center}
\label{fig:ex}
\caption{(color online). Numerical solutions of the Quantum Master Equation. (a): Time evolution of the QME populations in the absence of coupling 
[$J_{12}=0$, Eq.~(S2)]; (b): Time evolution of the QME populations in the presence of coupling [$J_{12}\neq\,0$, Eq.~(S3)].}
\end{figure}

The steady state solutions of the QME can be used to derive analytical expressions for the current generated in the reaction centre. 
In the case of non-interacting molecules ($J_{12}=0$) and for weak ambient pump conditions (i.e. $n_{1c}\,,n_{2c}\,,N_{c}\ll 1$)~\cite{mscully2013}, 
the current $\tilde{j}$ is given by 
\begin{equation}
\renewcommand{\theequation}{S\arabic{equation}} 
%\begin{split}
\label{eq:curr_non_int}
\displaystyle
\tilde{j}/e=\frac{\gamma_{c}\Gamma_{c}\gamma_{h}n_{h}}{\gamma_{c}\Gamma_{c}+ (\gamma_{c}+3\Gamma_{c})\gamma_{h}n_{h} 
+ \Gamma_{c}\gamma_{h} + (n_{h}\gamma_{c}\gamma_{h}\Gamma_{c})/\Gamma}\,.
%\end{split}
\end{equation}
When the donor molecules are coherently coupled ($J_{12}\neq0$),  and for the same conditions of weak ambient pump, we find the current $j$ to be
%\begin{widetext}
\begin{equation}
\renewcommand{\theequation}{S\arabic{equation}} 
\label{eq:curr_int}
%j/e = \left\{
%\begin{array}{ll}
\displaystyle
 j/e= \frac{n_{h}(1+n_{x})\gamma_{c}\Gamma_{c}\gamma_{h}\gamma_{x}}
 {[n_{h}(1+3n_{x}) + n_{x}]\Gamma_{c}\gamma_{h}\gamma_{x}+\gamma_{c}[(1+2n_{h})\Gamma_{c}\gamma_{h}+(1+n_{x})(\Gamma_{c}+n_{h}\gamma_{h})\gamma_{x}]
 +[n_{h}(1+n_{x})\gamma_{c}\Gamma_{c}\gamma_{h}\gamma_{x}]/\Gamma}\,.
\end{equation}
%\end{widetext}
%
\subsection*{Current-voltage characteristic and power at different temperatures}%\begin{figure}[h!]
As explained in the main text, for decreasing temperatures, the efficiency of charge transfer induced by the coherent coupling between the 
donor molecules can be substantially enhanced, thus leading to improved photocell performances, characterised by higher 
peak power. We can quantify the enhancement of the delivered peak power in the presence of coherent coupling, 
by defining the relative efficiency $\eta_{R}=(P_{out}^{\textrm{max}}-\widetilde{P}_{out}^{\textrm{max}})/\widetilde{P}_{out}^{\textrm{max}}$, 
$P_{out}^{\textrm{max}}$ and $\widetilde{P}_{out}^{\textrm{max}}$ being the output peak powers with ($J_{12}\neq\,0$) 
and without ($J_{12}=0$) coupling, respectively. For the parameters used in the simulation shown in Fig.~4, 
$\eta_{R}\approx$ 24\% at room temperature. The following figures show the current-voltage characteristic 
and power-voltage dependence for $T$=200K, 100K and 50K, evaluated for the same parameters of Fig.~4.
\begin{figure}[h!]
\renewcommand{\thefigure}{S\arabic{figure}}
\begin{center}
\subfigure{%
\label{fig:vj200}
\includegraphics[width=0.45\textwidth]{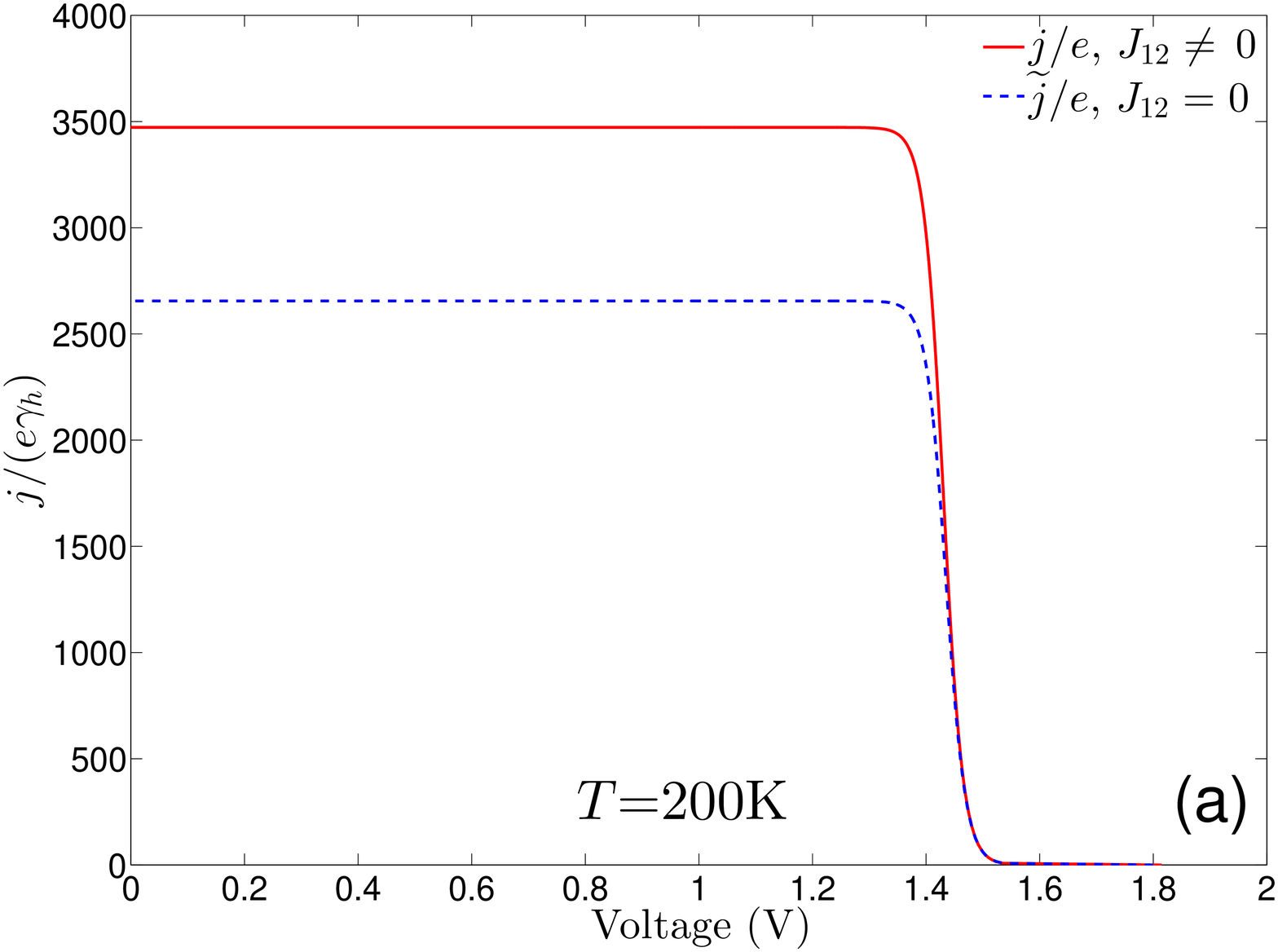}  
}%
\subfigure{%
\label{fig:vp200}
\includegraphics[width=0.45\textwidth]{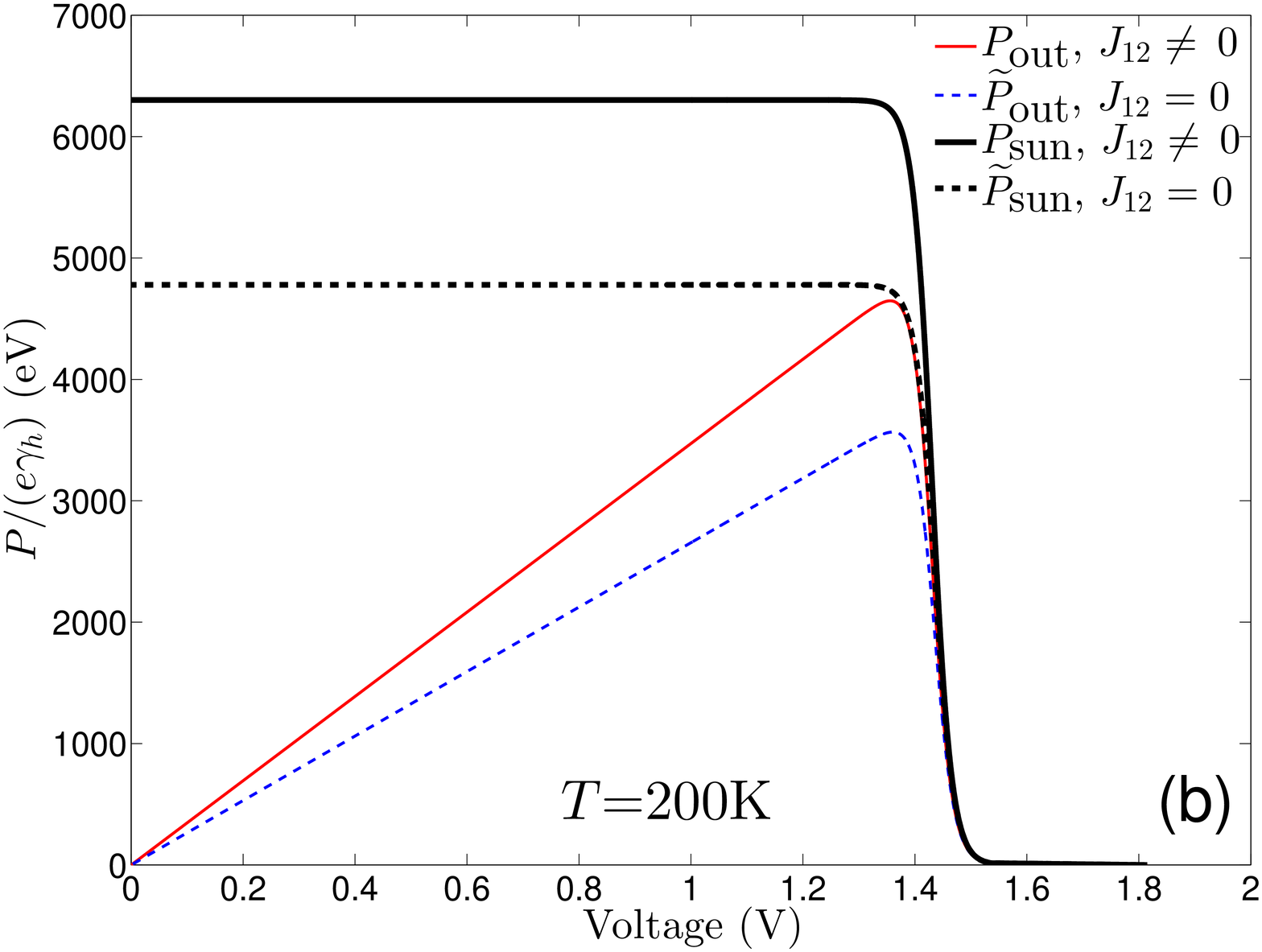} 
}\\%
\subfigure{%
\label{fig:vj100}
\includegraphics[width=0.45\textwidth]{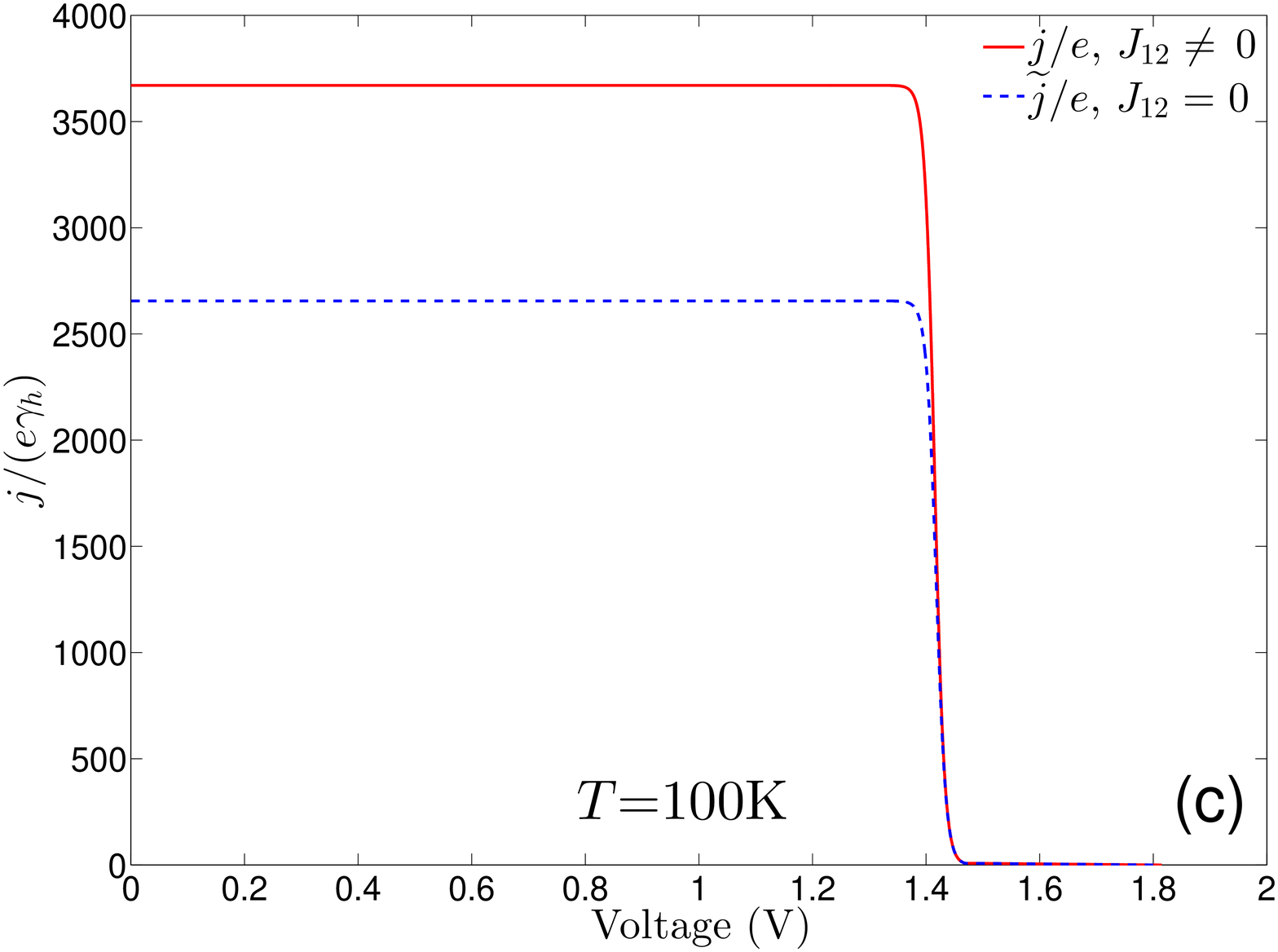}  
}%
\subfigure{%
\label{fig:vp100}
\includegraphics[width=0.45\textwidth]{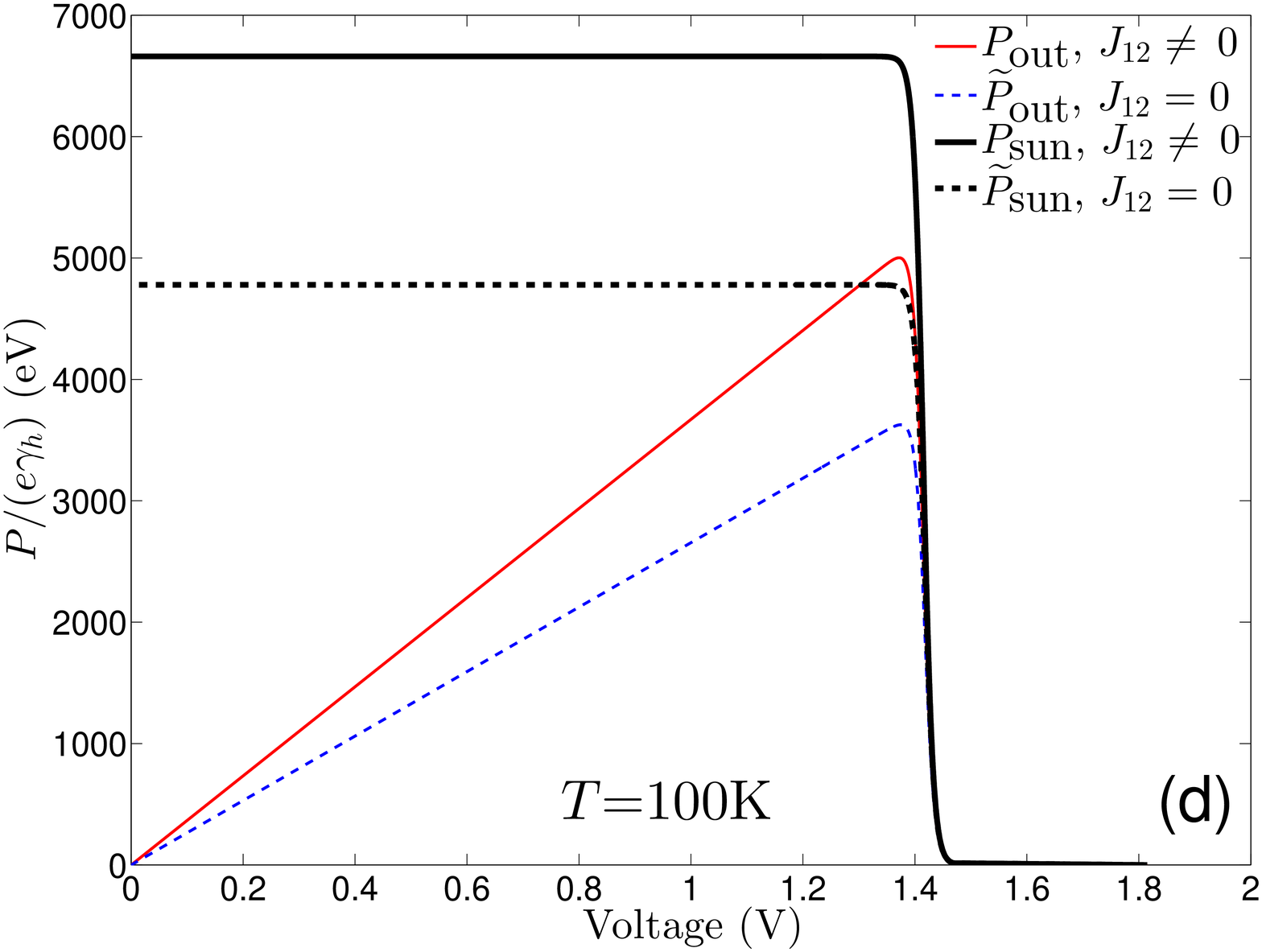}  
}\\%
\subfigure{%
\label{fig:vj50}
\includegraphics[width=0.45\textwidth]{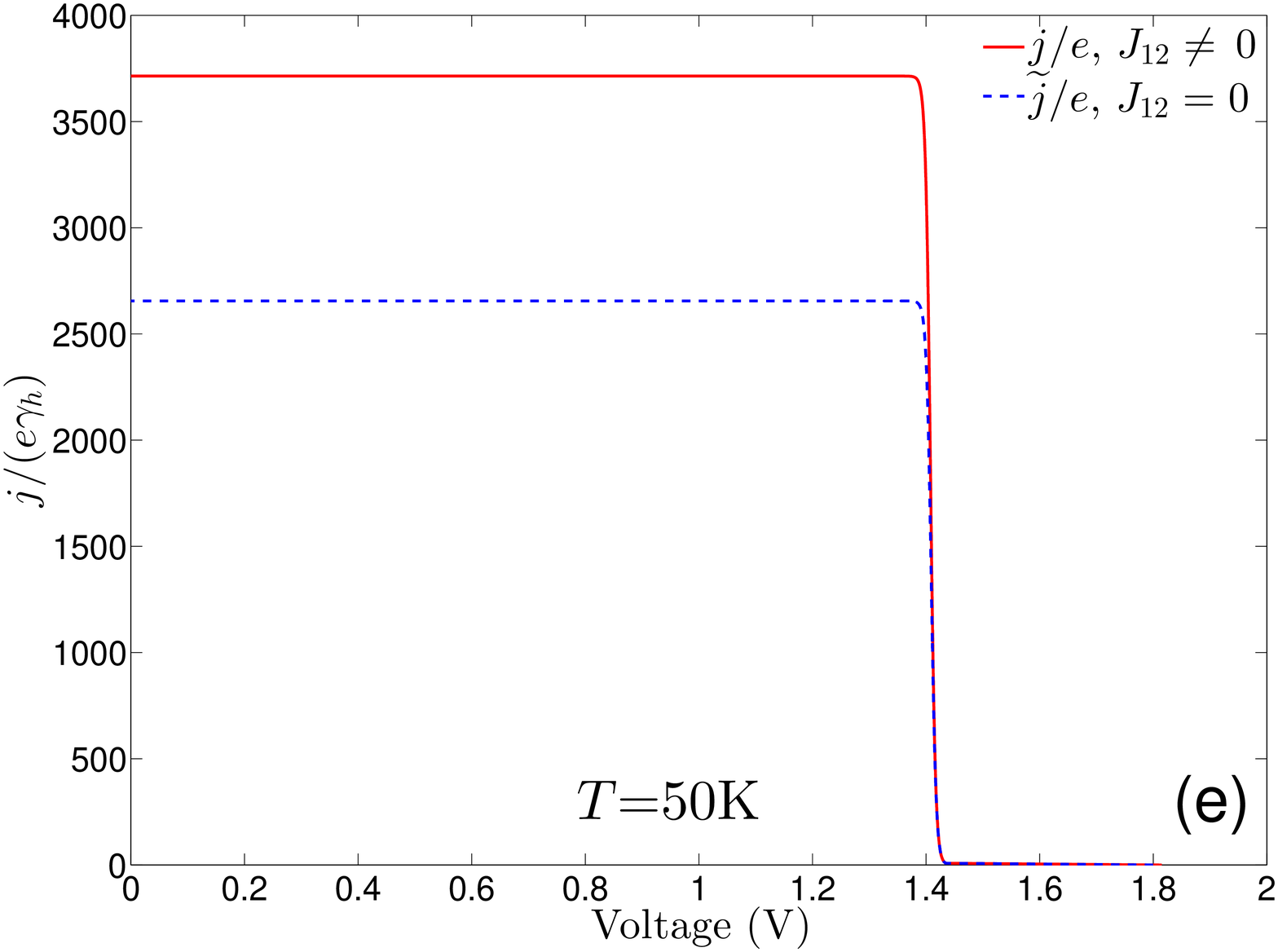} 
}%
\subfigure{%
\label{fig:vp50}
\includegraphics[width=0.45\textwidth]{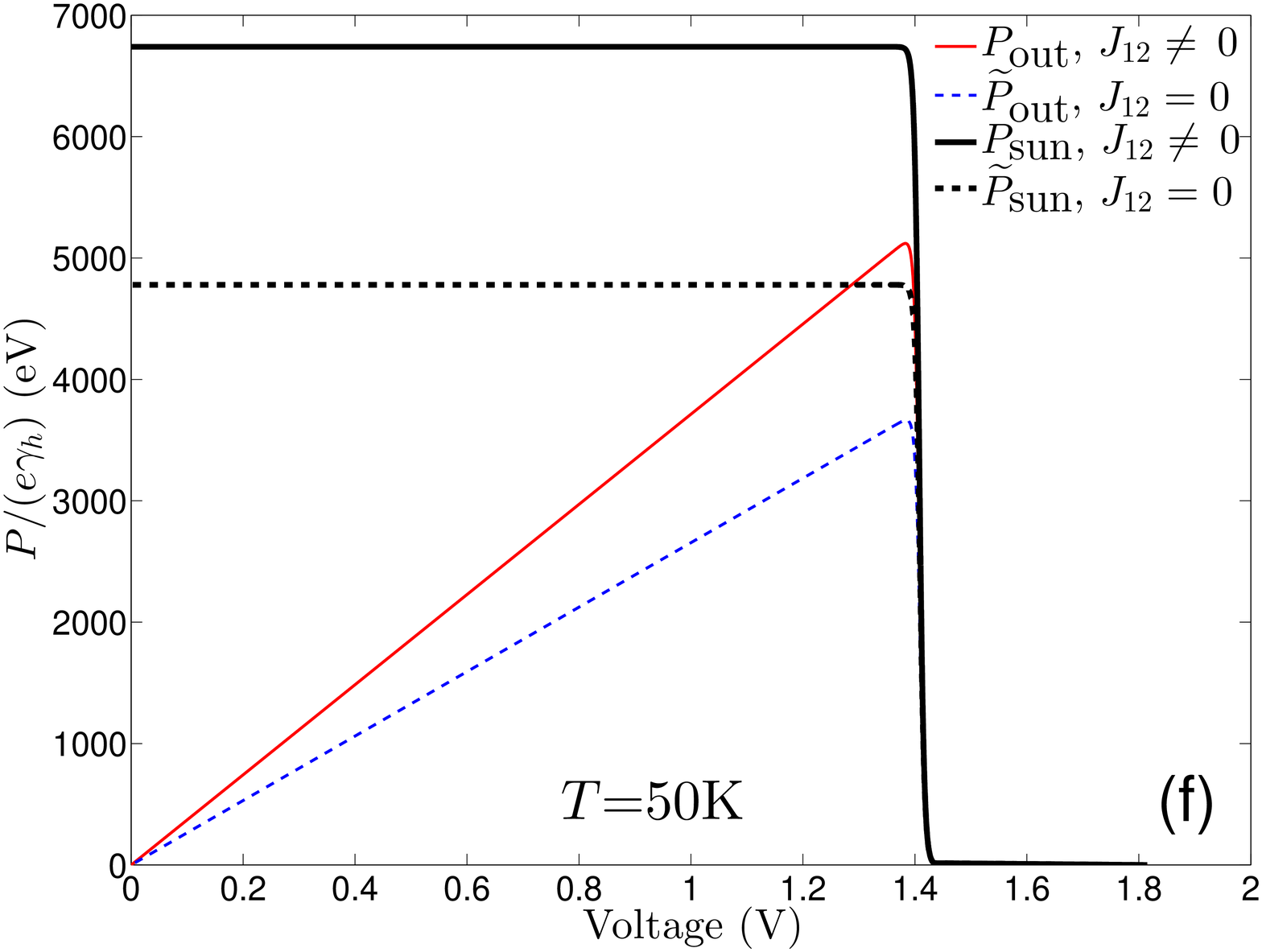} 
}%
\end{center}
\label{vjvp}
\caption{(color online). Current-voltage characteristic and power-voltage dependance at different ambient temperatures. (a)-(b): 
$T$=200K and $\eta_{R}\approx$ 30\%; (c)-(d): $T$=100K and $\eta_{R}\approx$ 38\%; (e)-(f): $T$=50K and $\eta_{R}\approx$ 40\%.}
\end{figure}

\clearpage

\subsection*{Comparison to previous work}%\begin{figure}[h!]
Dorfman {\it et al.} \cite{mscully2013} have recently proposed a model of a QHE based 
on a scheme similar to that studied in this paper, consisting of two donor molecules, an acceptor molecule and a cycling state. 
Figure 1(b) in the main text can be taken as a reference for the 
energy level scheme employed by Dorfman {\it et al.}. In [\onlinecite{mscully2013}], 
the coherent interaction between the excited donor states [$|a_{1}\rangle$ and $|a_{2}\rangle$ in Fig. 1(b)] 
is induced through the indirect process of Fano-interference, which results from 
coupling the excited states to a common `hot'  electromagnetic and `cold' phonon environment. 
The resulting dynamics of this cycle is rather more complicated than the one proposed in this paper, 
and requires the inclusion of coherence (off-diagonal) terms in the density matrix evolution. 
The Fano interference induces population-to-coherence transfer, a phenomenon that also appears 
in non-secular Bloch-Redfield theory (BRT); the Fano master equation can also be derived by applying the 
non-secular BRT to the light-matter interaction \cite{mscully2013}. 
However, the inclusion of non-secular terms in BRT master equations is known to permit unphysical solutions for 
all system-reservoir interactions, other than pure dephasing~\cite{superconductors}. These unphysical solutions appear as density matrices 
with negative eigenvalues, which can also manifest as negative populations for some states. Furthermore, such violations are 
expected to increase with the strength of the non-secular interactions~\cite{superconductors}. 

We have solved the quantum master equation reported in [\onlinecite{mscully2013}] (see Eqs.~S74-S79 in [\onlinecite{mscully2013}]) using the 
initial conditions and parameters specified in the paper. Figures S3(a)-S3(c) illustrate the time-evolution of both the populations of $|a_{2}\rangle$ 
and $|b\rangle$ ($\rho_{22}\equiv\rho_{a_{2}a_{2}}$ and $\rho_{bb}$) as well as the coherence between $|a_{1}\rangle$ and $|a_{2}\rangle$ 
($\rho_{12}$) in the `overdamped' regime. Figures S3(a) and S3(c) are in perfect agreement with those shown by Dorfman {\it et al.} 
(see Figs.~3C-3D in [\onlinecite{mscully2013}]). Figure S3(b) has been obtained calculating the dynamics at times larger than those shown in [\onlinecite{mscully2013}] 
and clearly shows that the `overdamped' regime, in which Dorfman {\it et al.} report the largest enhancement of photocurrent, 
is characterised by a steady state with a large, negative population in the $|a_{2}\rangle$ state. This regime corresponds to the limit where the 
driving of the coherence due to coherence-to-population transfer terms is strong, and comparable to the free oscillation frequency of the coherence. 
Interestingly, for intermediate coupling a more modest enhancement is found, but the steady state, which also possesses stead
oscillatory coherence [see Fig. S3(e)], does remain positive. 

Moreover, on going deeper into this regime, reducing the energy 
difference between the donor energy levels, we find that the master equation reported in [\onlinecite{mscully2013}] is unstable 
and the individual populations may diverge, although their sum always remains unity due to the negative populations. 
This is exemplified by Fig. S4, which shows the steady-state solutions for the $|a_{2}\rangle$ and $|\alpha\rangle$ populations 
as a function of the alignement factor of the photon dipole matrix element $p_{h}$ (see e.g. [\onlinecite{mscully2011b}]). 
At maximum interference, the  Fano-induced photonic coupling rate is given by $\gamma_{12h}=p_{h}\sqrt{\gamma_{1h}\gamma_{2h}}$ 
with $p_{h}=1$. This is a rather subtle point of the model proposed by Dorfman {\it et al.}: in [\onlinecite{mscully2013}] the authors 
do not introduce the alignment factor within their model and equations, and thus we assume that $p_{h}=1$; in 
other publications (see e.g. [\onlinecite{mscully2011b}]) the same authors use $p_{h}=1-\varepsilon$, $\varepsilon$ being a small number, without any 
further clarification. However, Fig. S4 clearly shows that as the energy spacing between the excited states ($a_{1}$ and $a_{2}$) decreases, the density matrix 
element $\rho_{22}$ becomes negative for a wide range of values of the alignment factor $p_{h}$ and both $\rho_{22}$ and $\rho_{\alpha\alpha}$ 
tend to diverge.
 
In light of this, it is not clear if the enhancement of photocurrent induced by Fano interference is a real feature of the proposed model or an artefact of the 
master equation used to simulate its dynamics (which also neglects incoherent relaxation between the optical excited states). 
Indeed, as pointed out by Nalbach and Thorwart in their commentary on [\onlinecite{mscully2013}], it is perhaps surprising that the enhancements 
of photocurrent and steady-state coherence oscillations only appear in the strongly damped regime~\cite{nalbach2013}.  
 All this suggests that this mechanism may lead to enhancement (none is seen for weak coupling), but a more sophisticated, 
 non-perturbative method of simulating the dynamics - starting with the microscopic interactions - 
 should be used to assess such a claim. We note that recent numerically exact studies of exciton dynamics in PPCs have shown that population to coherence dynamics \emph{can} arise when the phonon environment contains contributions from non-adiabatic coupling to underdamped vibrations~\cite{achin2013}. 
 
 It is also worth noting that in [\onlinecite{mscully2013}], the QME has been solved assuming the system initially being in the coherent superposition state, i.e. 
using $\rho_{11}(t=0)=\rho_{22}(0)=\rho_{12}(0)=0.5$. These conditions do not characterise a system under incoherent excitation by incident sunlight, which should be 
described by a initially fully occupied ground state , i.e. $\rho_{bb}(t=0)=1$. We have solved the QME given in [\onlinecite{mscully2013}] using this latter initial condition, 
and found that even in this case, the population of the $|a_{2}\rangle$ state becomes negative almost immediately [see Fig.~S5(a)]. Moreover, this pathology can be observed 
also in other regimes investigated by Dorfman {\it et al.}. Figure S5(c), for instance, shows that negative values of the populations are obtained when solving the QME of [\onlinecite{mscully2013}] in the `intermediate regime' under the conditions of incoherent excitation [$\rho_{bb}(t=0)=1$].

\begin{figure}[h!]
\renewcommand{\thefigure}{S\arabic{figure}}
\begin{center}
\subfigure{%
\label{fig:scully1}
\includegraphics[width=0.45\textwidth]{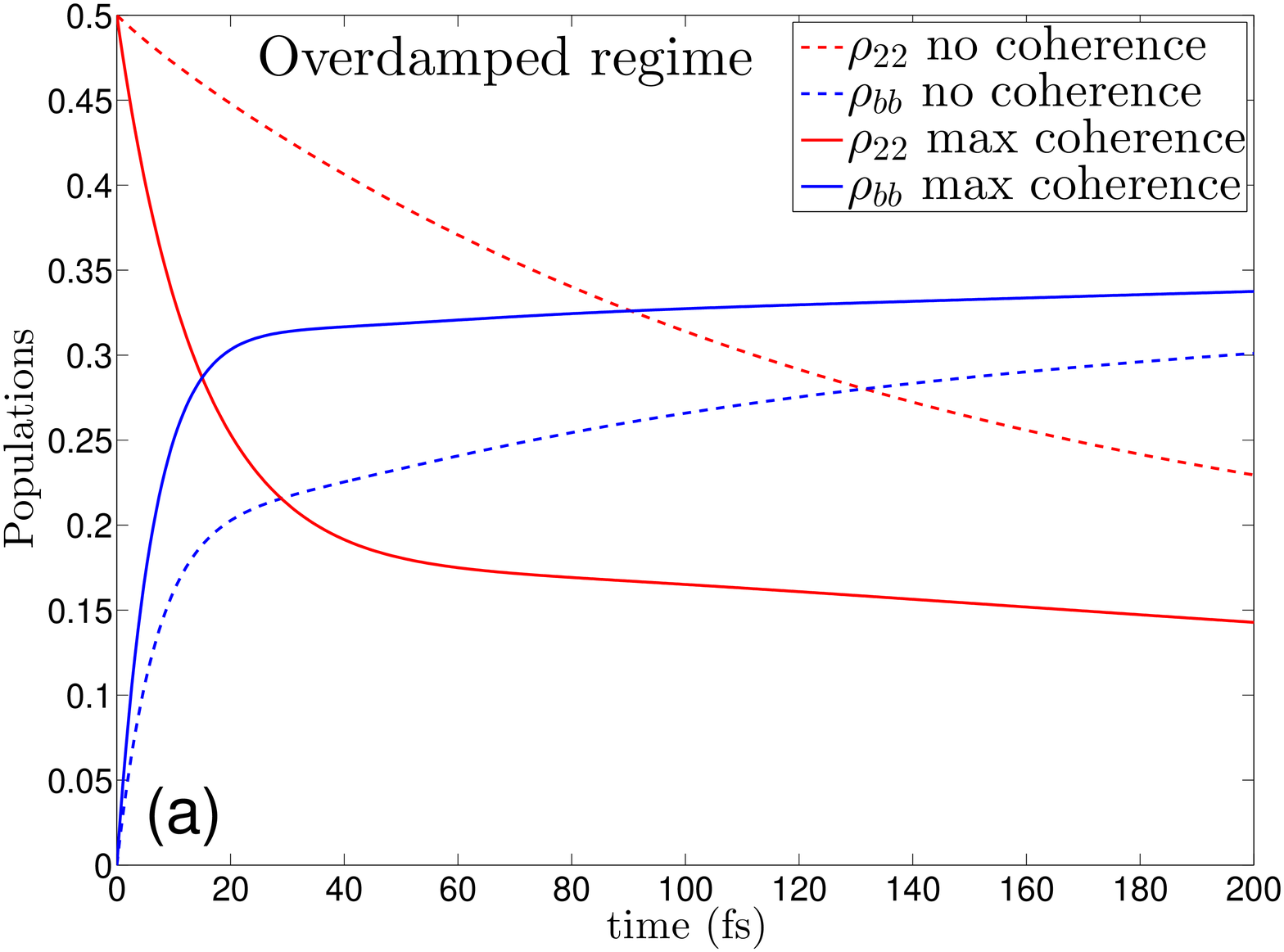}  
}%
\subfigure{%
\label{fig:scully2}
\includegraphics[width=0.45\textwidth]{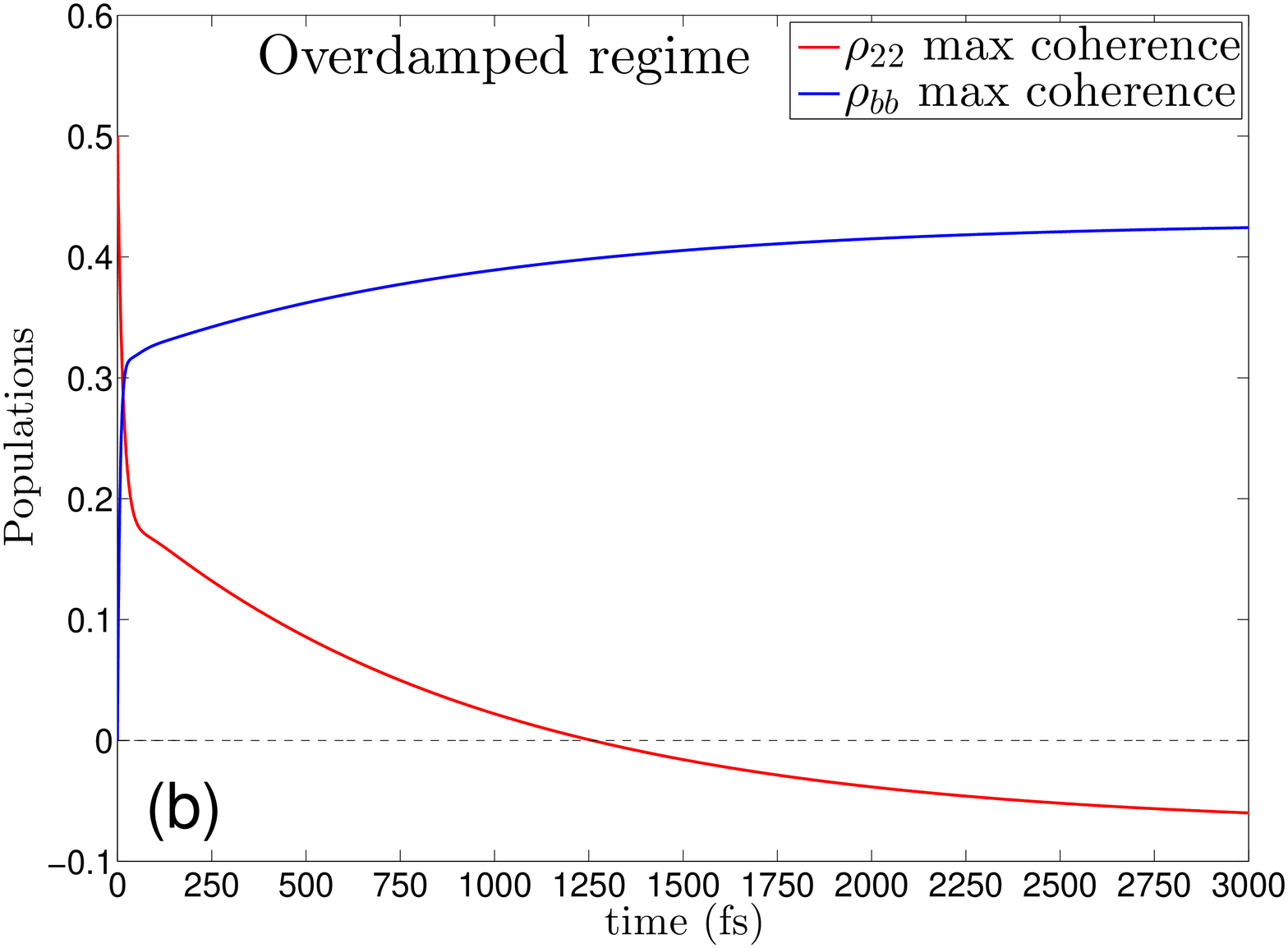}  
}\\%
\subfigure{%
\label{fig:scully3}
\includegraphics[width=0.45\textwidth]{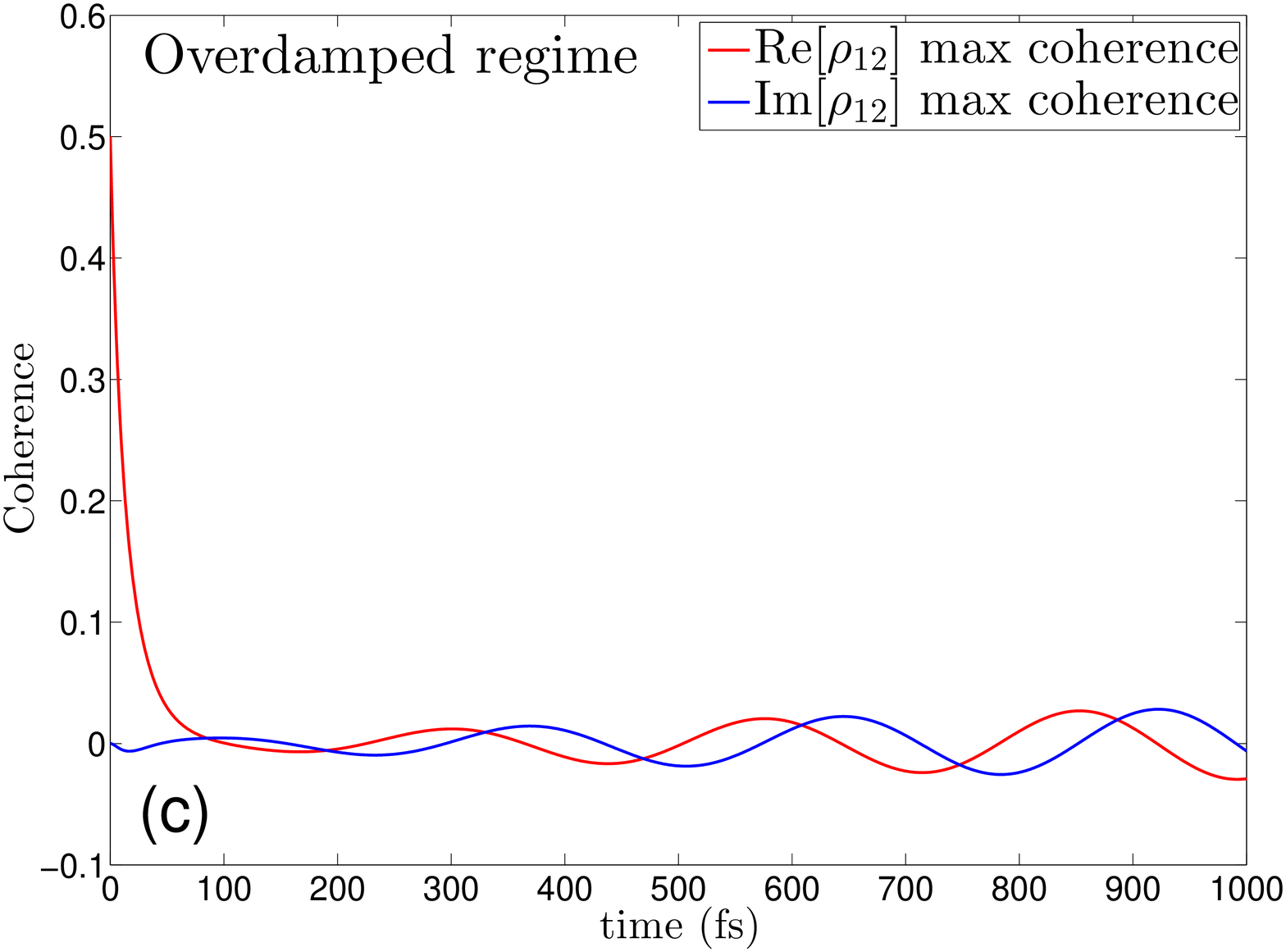}  
}%
\subfigure{%
\label{fig:scully4}
\includegraphics[width=0.45\textwidth]{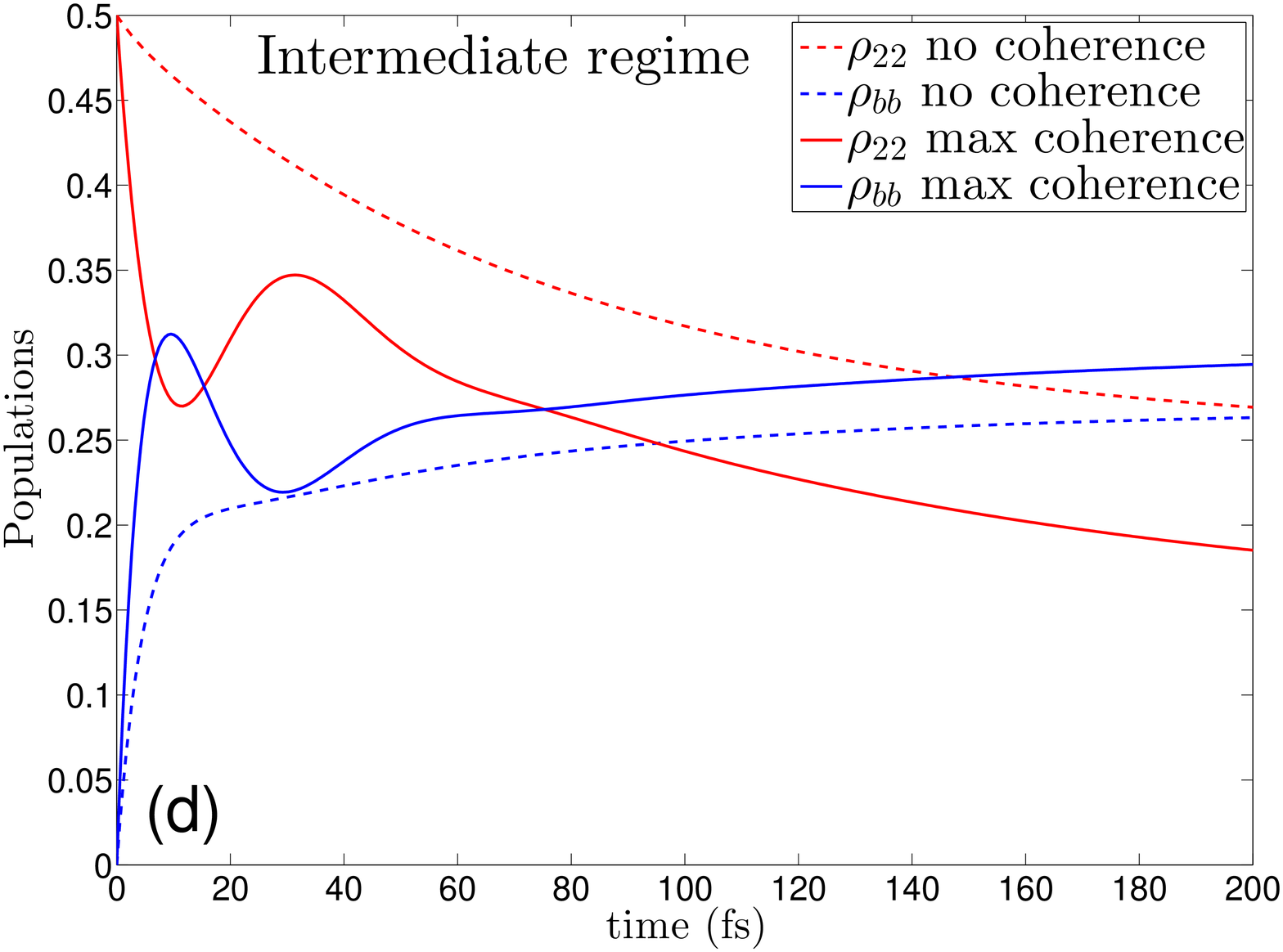}  
}\\%
\subfigure{%
\label{fig:scully5}
\includegraphics[width=0.45\textwidth]{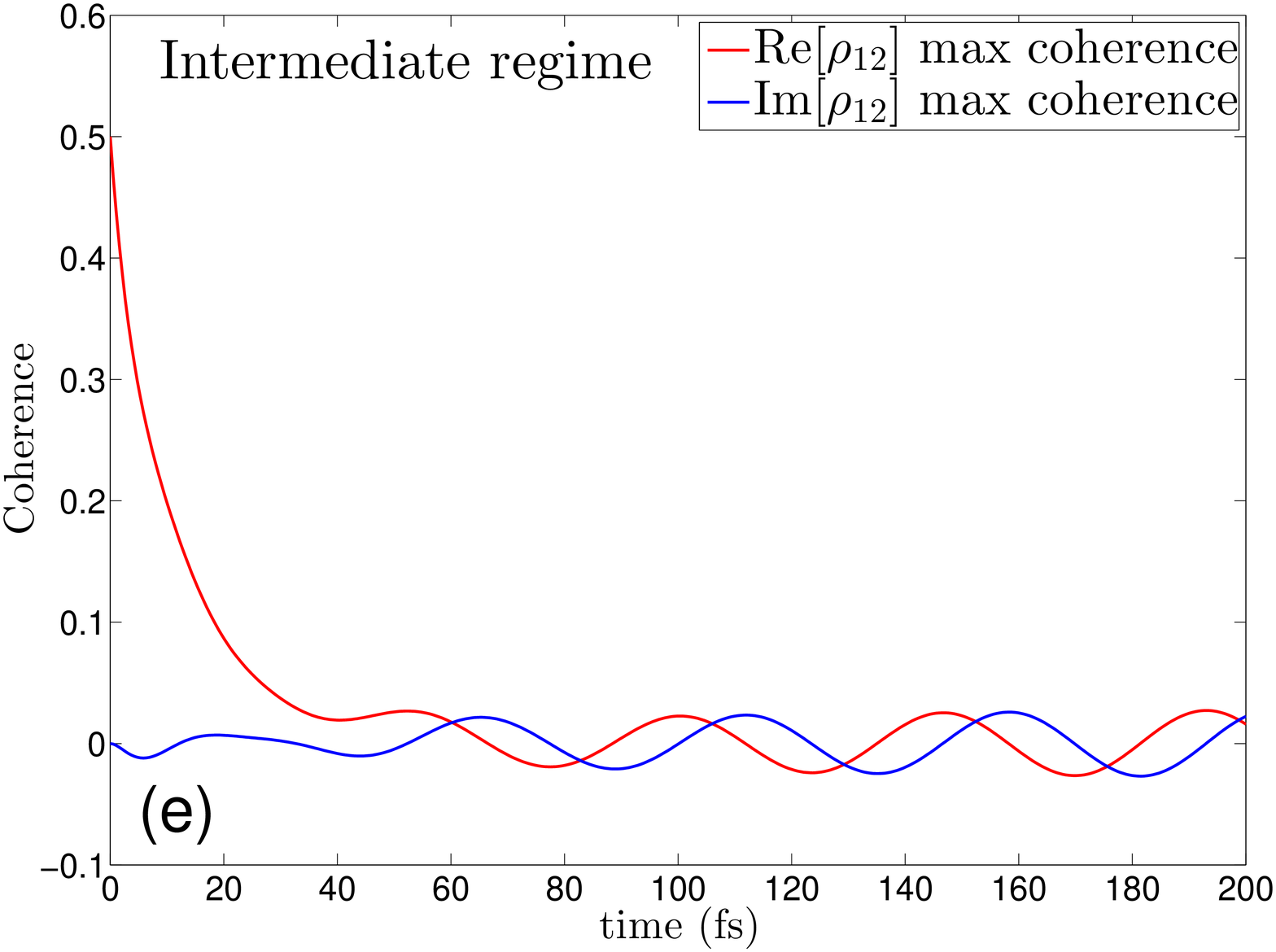}  
}%
\end{center}
\label{scully1}
\caption{(color online). Numerical solutions of the model proposed by Dorfman {\it et al.} 
(a)-(c): Time-evolution of the populations of states $|a_{2}\rangle$ and $|b\rangle$ and coherence ($\rho_{12}$) in the `overdamped regime', 
see also Fig.~3C-3D in [\onlinecite{mscully2013}]. (d)-(e): Time-evolution of the populations of states $|a_{2}\rangle$ and $|b\rangle$ 
and coherence ($\rho_{12}$) in the `intermediate regime', see also Fig.~3K-3L in  [\onlinecite{mscully2013}].} 
\end{figure}

\begin{figure}[h!]
\renewcommand{\thefigure}{S\arabic{figure}}
\centering
\includegraphics[width=0.6\textwidth]{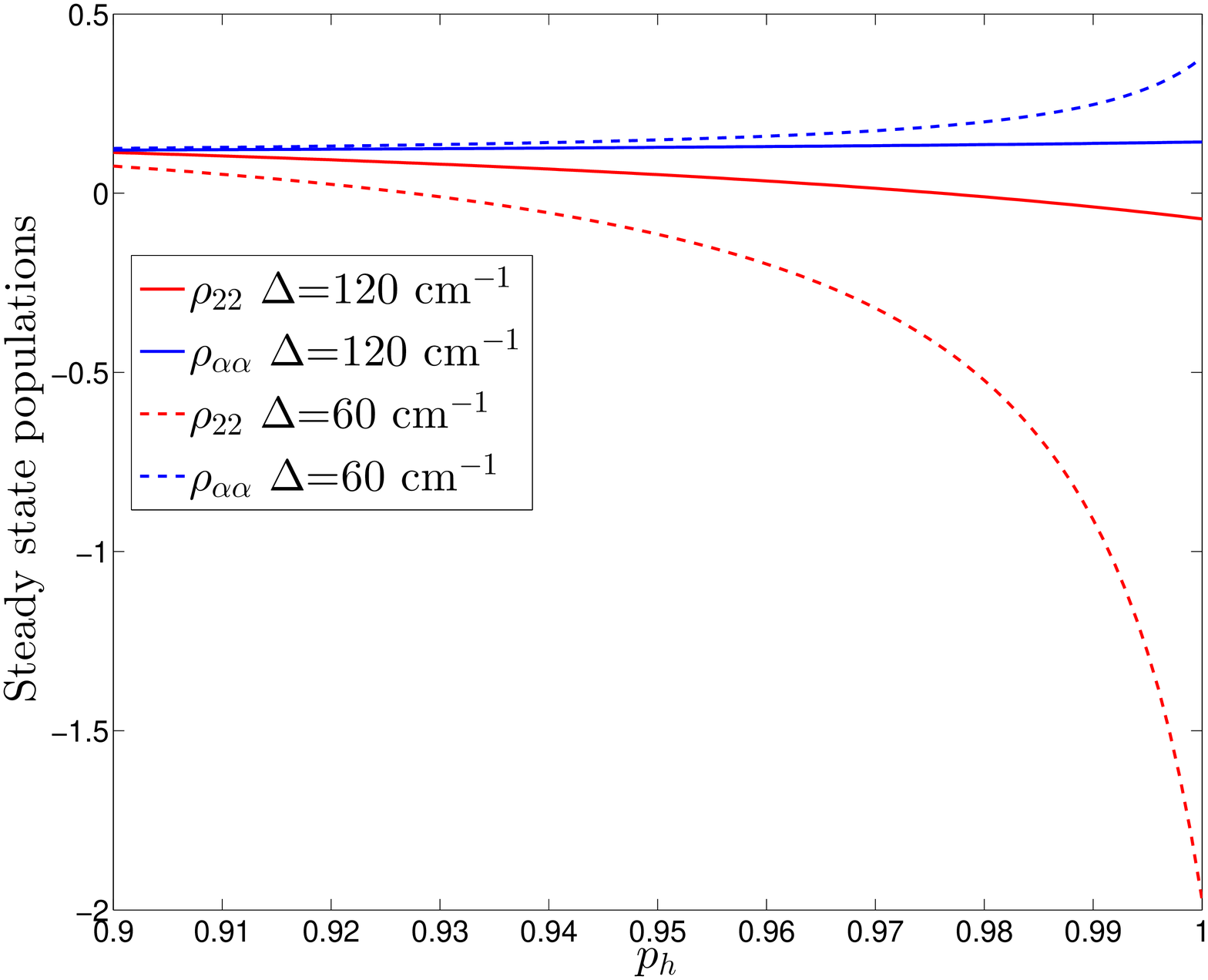}%model.eps
\caption{(color online). Numerical solutions of the model proposed by Dorfman {\it et al.} 
Time-evolution of the populations of states $|a_{2}\rangle$ and $|\alpha\rangle$ in the `overdamped regime', 
as a function of the photonic alignment factor $p_{h}$ and for energy detuning between the molecular excites states ($a_{1}$, $a_{2}$) 
$\Delta=120$ cm$^{-1}$ as in Fig.~3 in [\onlinecite{mscully2013}] and $\Delta=60$ cm$^{-1}$.}
\label{fig:scully2}
\end{figure}

\begin{figure}[h!]
\renewcommand{\thefigure}{S\arabic{figure}}
\begin{center}
\subfigure{%
\label{fig:scully1}
\includegraphics[width=0.33\textwidth]{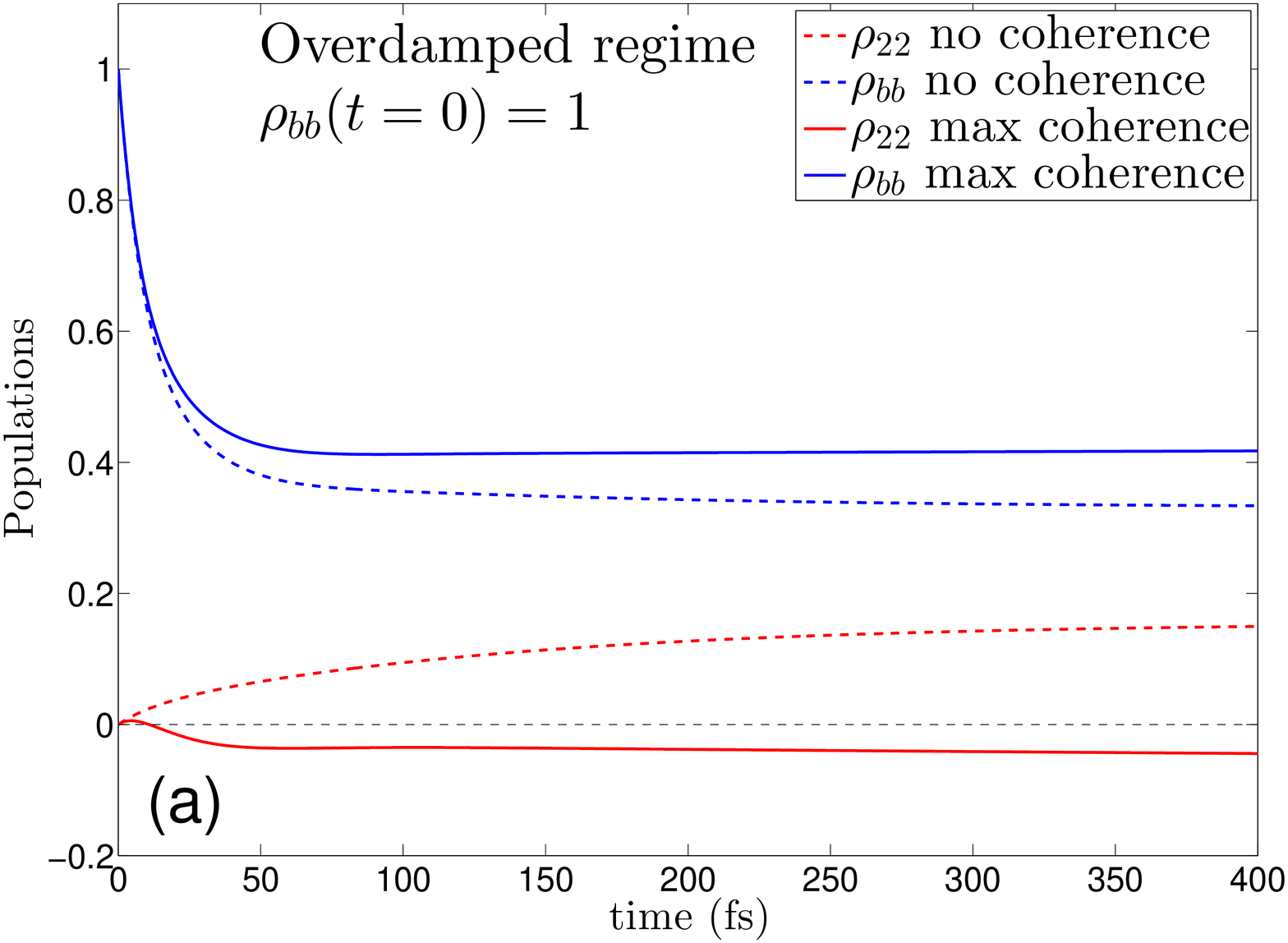}  
}%
\subfigure{%
\label{fig:scully2}
\includegraphics[width=0.33\textwidth]{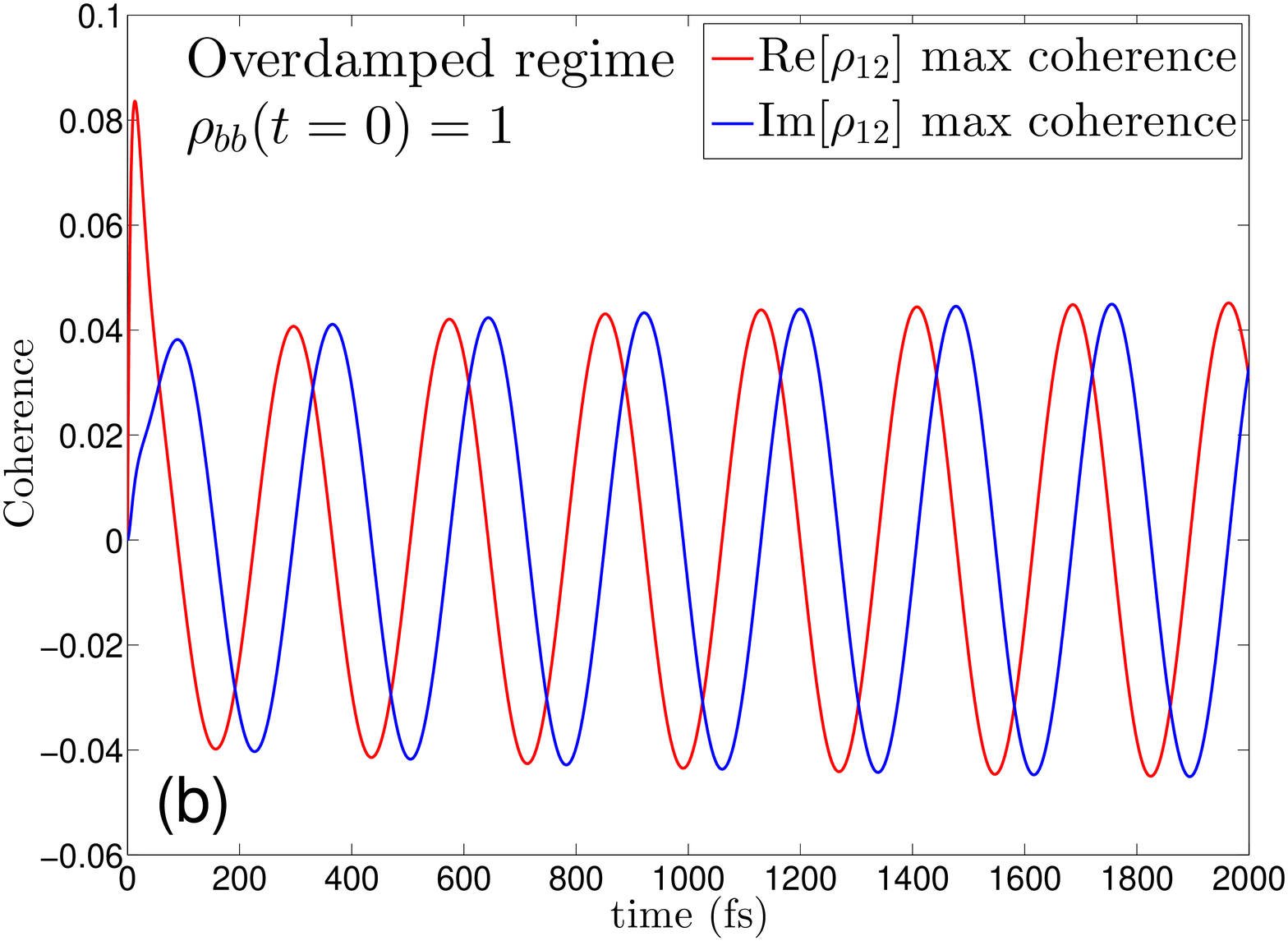}  
}\\%
\subfigure{%
\label{fig:scully3}
\includegraphics[width=0.33\textwidth]{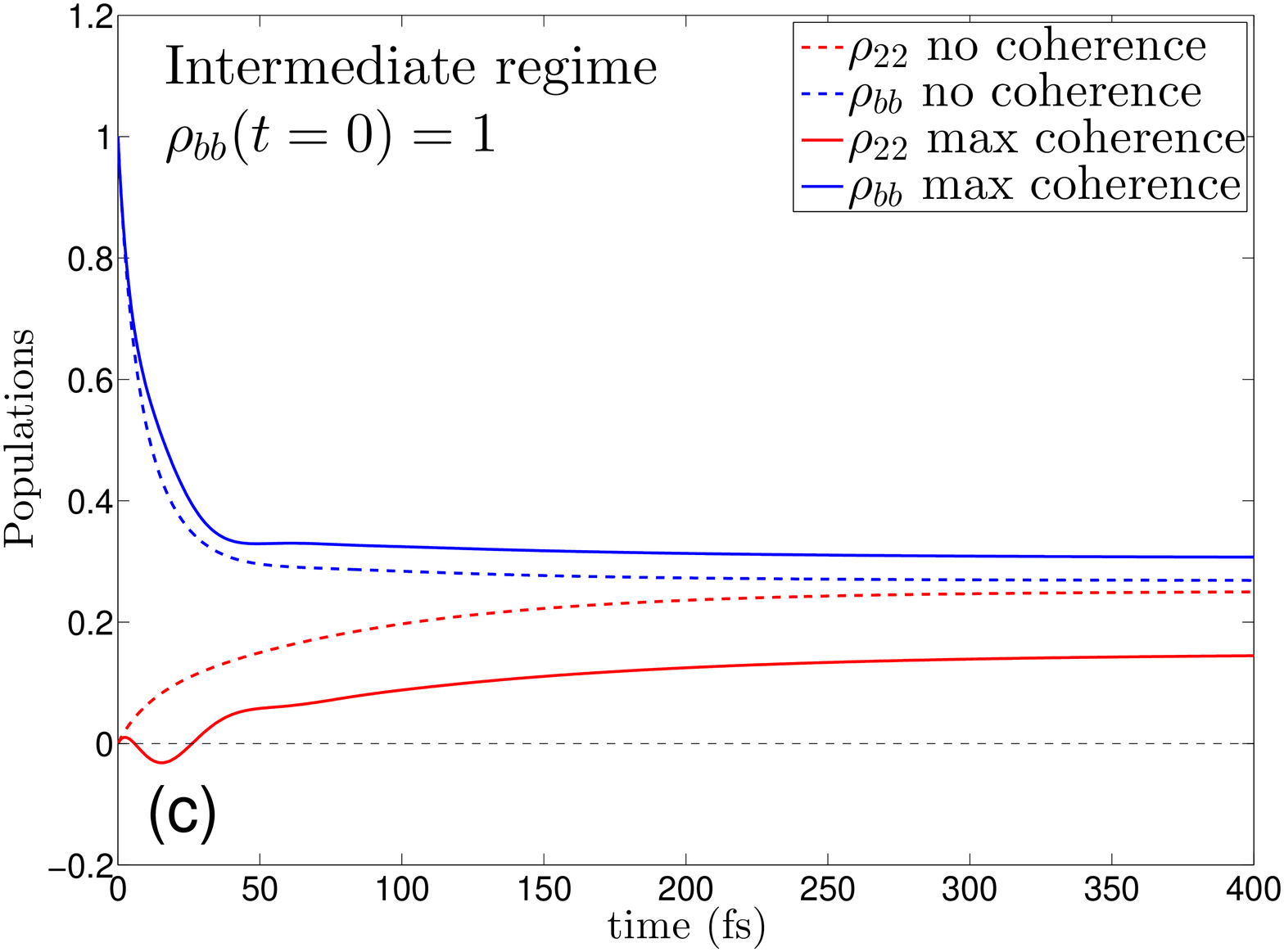} 
}%
\subfigure{%
\label{fig:scully4}
\includegraphics[width=0.33\textwidth]{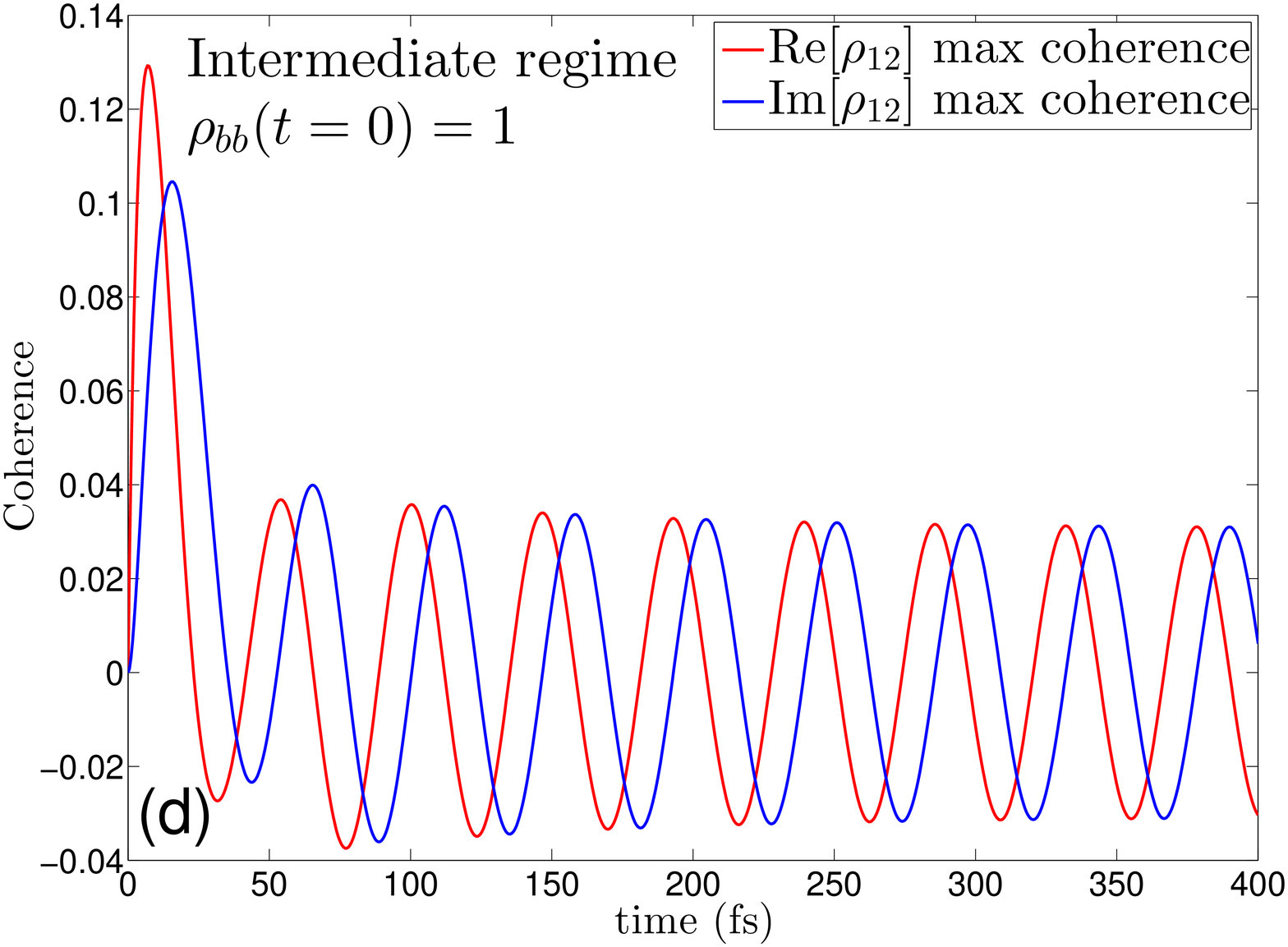}  
}%
\end{center}
\label{scully1}
\caption{(color online). Numerical solutions of the model proposed by Dorfman {\it et al.} 
Time-evolution of the populations of states $|a_{2}\rangle$ and $|b\rangle$ and coherence ($\rho_{12}$) in the `overdamped regime' (a)-(b) and 
in the `intermediate regime' (c)-(d), obtained solving the QME given in  [\onlinecite{mscully2013}] using $\rho_{bb}(t=0)=1$ as initial condition.} 
\end{figure}

\end{document}